\documentclass[11pt,twosided]{article}
                                           \usepackage{amsmath}
\usepackage{color}
\usepackage{amsfonts}
\usepackage{amssymb}
\usepackage{bbold}
\usepackage{graphicx}
\usepackage{verbatim}
\numberwithin{equation}{section}
\usepackage{array}
\date{}
\title{\textbf{Shift symmetries and duality web in gauge theories}}
\begin{document}
	\maketitle
		\centerline{Rabin Banerjee{\footnote{Raja Ramanna Fellow; rabin@bose.res.in}} and Anwesha  Chakraborty{\footnote{anwesha@bose.res.in}}}
	\centerline{S.N. Bose National Centre for Basic Sciences, Kolkata- 700106, India}
	\begin{abstract}
		Using a generalised Noether prescription we are able to extract all the currents and their conservation laws in space dependent shift symmetric theories. Various identities among the currents in the matter sector are found that form the basis for revealing a dual picture when the full interacting theory is considered by coupling to gauge fields. The coupling is achieved in terms of vector fields by adhering to a modified minimal prescription which is also supported by an iterative Noether scheme. Further, this scheme shows that couplings can also be introduced using higher rank tensor gauge fields that have appeared in recent discussions on fractons. We reveal a connection among these descriptions (using vector or tensor fields) through certain duality maps that relate the various fields (gauge, electric and magnetic) in the two cases. A  correspondence is established among the Gauss' law, Faraday's law and Ampere's law. Explicit calculations are provided for  linear and quadratic shift symmetric lagrangians. 
		
	\end{abstract}
	\tableofcontents
	\section{Introduction }
Recent studies on quantum phases of matter have indicated the existence of peculiar quasiparticles, called fractons, with the characteristic feature of restricted mobility (for reviews, see \cite{nandkishore,pretko2020,bennett,hoyos}). Since this is a novel (theoretical) phase of matter, it has triggered intense activity covering a wide gamut of topics like field theories with higher rank symmetric tensors \cite{pretko2020,jensen,slagle,nathan,argurio,seiberg,rb,dual} gravitational physics \cite{rb,xu,pretko2,moi,kurt}, holography \cite{yan} and even quantum information storage \cite{haah,haah2,terhal,brown}.\\\\
The preeminent defining property of fractons, namely, their partial or complete immobility \cite{chamon,vijay,vijay2,blasi} is attained by invoking a rank-2 spatial symmetric tensor gauge field $A_{ij}$ instead of the usual vector field $A_i$. To  visualise this it may be recalled that the vector theory with standard gauge symmetry,
\begin{equation}
    \delta A_i=\partial_i\Lambda\label{i1}
\end{equation}
and Gauss constraint,
\begin{equation}
    \partial_iE_i=\rho
\end{equation}
where $E_i$, the electric field, is the momentum conjugate to $A_i$ and $\rho$ is the charge density,
admits two possibilities for extending to symmetric tensor fields,\\
(a) as a vector charge theory with,
\begin{align*}
    \delta A_{ij}&=\partial_i\Lambda_j+\partial_j\Lambda_i\\
    \partial_i E^{ij}&=\rho^j
\end{align*}
where $E_{ij}$, the generalised electric field, is the momentum conjugate to $A_{ij}$ and $\rho^j$ is the vector charge density, or, alternatively,\\
(b) as a scalar charge theory with,
\begin{align}
    \delta A_{ij}&=\partial_i\partial_j\Lambda\nonumber\\
    \partial_i\partial_jE^{ij}&=\rho\label{i2}
\end{align}
Both possibilities impose restrictions on the fracton mobility which is partial for case (a) but complete for case (b) \cite{pretko3}. This is obvious from (\ref{i2}) since it implies simultaneous conservation of charge as well as dipole moment,
\begin{equation}
    \int_{space}\,\rho=\int_{space}\, x^k\rho=0
\end{equation}
Also, comparing gauge transformation properties given in (\ref{i1}) and (\ref{i2}) suggests that the tensor and vector fields are related by,
\begin{equation}
    A_{ij}=\frac{1}{2}(\partial_iA_j+\partial_jA_i)\label{i4}
\end{equation}
In this paper we focus on the scalar charge theory defined in case (b) and we show that, using \eqref{i4} it is possible to have another description in terms of a vector field $A_i$ with conventional transformation property (\ref{i1}). 
The full interaction theory is considered where the starting point is the matter sector defined in terms of higher derivatives of complex scalars. While the explicit form of the matter sector is known \cite{dual,pretko4,noether,yau,wang2} it is derived here from a Lifschitz type action involving real scalars with polynomial shift symmetries. Gauge couplings are now introduced in the matter sector by following the iterative Noether's prescription of Deser \cite{deser}. Two variants are revealed. Interactions may be introduced either by a vector field $A_i$ or a symmetric rank-2 field $A_{ij}$, which are related by (\ref{i4}). The results for the vector coupling may, alternatively, be understood as following from a generalised minimal prescription applied to the matter sector. The tensor coupling, on the other hand, exactly corresponds to the gauging advocated by the fracton gauge principle \cite{dual,pretko4}. Once the matter sector and interactions have been obtained, it remains to define the pure gauge sector. For the fractonic theory this is known in terms of $A_{ij}$. Its alternative description is obtained by using the map (\ref{i4}). We thus have two clear choices for the fractonic theory expressed either in terms of $A_{ij}$ or $A_i$. A full dynamical treatment, either in the lagrangian or hamiltonian formulations, reveals a correspondence among the equations of motion, Gauss constraint, Faraday's law and Ampere's law. We also discuss some contrasting features of the two descriptions. We extend this correspondence further by considering conservation of the quadrupole moment, besides charge and dipole moment. Now there is a duality web involving $A_i,A_{ij}$ and $A_{ijk}$ where the rank-3 tensor appears due to the new conservation of quadrupole moment.\\\\
The paper is structured as follows: in section 2 we consider arbitrary shift symmetric real scalar and vector theories and derive identities among various conserved currents. Alternative forms for these currents are obtained that differ by local counterterms. By suitable maps, shift symmetries in real scalar theories are identified with gauge symmetries in complex scalar theories in section 3. These maps enable us to determine the matter sector as well as expressions for conserved currents found in the literature \cite{seiberg,dual,pretko4,noether} on fractons. Sections 4 and 5, respectively, deal with dual formulations in linear and quadratic shift symmetric theories in 2+1 and 3+1 dimensions, employing both lagrangian and hamiltonian techniques. Finally, conclusions are given in section 6.  
	\section{Shift Symmetries and conservation laws}
	An important input in the study of low-energy effective lagrangians is Goldstone's theorem which implies that whenever a global symmetry is spontaneously broken, a gapless mode will appear. For relativistic theories, this leads to a massless Goldstone particle. Such a particle is described by a shift symmetry of the field, in the example of a Goldstone scalar,
	\begin{equation}
		\phi(x)\,\to\,\phi(x)+c\label{k1}
	\end{equation}
	where $c$ is a constant, and is characterised by the usual action of a scalar field,
	\begin{equation}
		S=\frac{1}{2}\int\,\,d^dx\,\,\partial_{\mu}\phi\,\partial^{\mu}\phi\label{k2}
	\end{equation}
	Since the action is invariant under (\ref{k1}), which is a global transformation, we may obtain the corresponding conserved current by using Noether's first theorem,
	\begin{equation}
		j^{\mu}=\frac{\partial\mathcal{L}}{\partial(\partial_{\mu}\phi)}\delta\phi=c\,\partial^{\mu}\phi\label{k3}
	\end{equation}
	which satisfies,
	\begin{equation}
		\partial_{\mu}j^{\mu}=c\,\,\Box \phi=0\label{k4}
	\end{equation}
	on using the equation of motion. Further, it is found that the shift symmetry, on exploiting canonical quantisation of $\phi$, connects the vacuum $|0\rangle$ with a single particle state $|1\rangle$,
	\begin{equation}
		\langle 0|j^{\mu}|1\rangle =c \langle 0|\partial^{\mu}\phi|1\rangle =cp^{\mu}e^{ip.x}\label{k5}
	\end{equation}
	and  the matrix element vanishes in the strict soft limit, as required by Goldstone's theorem.\\ \\
	We now generalise the above analysis by infusing two modifications. First, a non-relativistic version is taken and, secondly, the constant shift symmetry is enlarged to a linear spatial shift, 
	\begin{equation}
		\phi\,\to\, \phi+c+c_ix_i=\phi+w\label{k6}
	\end{equation}
	The non-relativistic free scalar theory is defined as,
	\begin{equation}
		\mathcal{L}=\frac{1}{2}\dot{\phi}^2-\frac{1}{2}(\partial_i\phi)(\partial_i\phi)=\frac{1}{2}\partial_0\phi\partial_0\phi-\frac{1}{2}\partial_i\phi\partial_i\phi\label{k7}
	\end{equation}
	which transforms in the following way,
	\begin{equation}
		\delta\mathcal{L}=-(\partial_i\phi)(\partial_iw)=-\partial_i(\phi\partial_i w)\label{k8}
	\end{equation}
	Thus the Lagrangian is quasi-invariant.\\\\
	Applying Noether's theorem, the conserved currents are given by,
	\begin{align}
		j^0&=\frac{\partial\mathcal{L}}{\partial(\partial_0\phi)}\delta\phi=(\partial_0\phi) \,w=\dot{\phi}\,w\label{k9a}\\
		j^i&=\frac{\partial\mathcal{L}}{\partial(\partial_i\phi)}\delta\phi+\phi\partial^iw=-(\partial^i\phi)\,w+\phi(\partial^i\,w)\label{k9b}
	\end{align}
	where the extra piece in (\ref{k9b}) is the boundary term in (\ref{k8}). Current conservation now follows on shell,
	\begin{equation}
		\partial_0j_0-\partial_ij_i=(-\Ddot{\phi}+\partial^2\phi)w=0\label{k10}
	\end{equation}
	where $\partial^2 \, w=0$ has been used since $w$ is linear in $x$ (\ref{k6}).\\\\
	However, there are other possibilities for defining the currents. A more conventional definition is,
	\begin{align}
		j_0&=-j^0=-\dot{\phi}\label{k11a}\\
		j_i&=j^i=-\partial_i\phi\label{k11b}
	\end{align}
	On shell current conservation follows immediately,
	\begin{equation}
		\partial_0j_0-\partial_ij_i=-\Ddot{\phi}+\partial^2\phi=0\label{k12}
	\end{equation}
	It is now possible to introduce a tensor current $j_{ki}$ whose space derivative yields the vector current,
	\begin{equation}
		j_{ki}=-\phi\delta_{ki}, \qquad j_i=\partial_kj_{ki}\label{k13}
	\end{equation}
	so that,
	\begin{equation}
		\partial_0j_0-\partial_i\partial_kj_{ki}=0\label{k14}
	\end{equation}
	The above relation shows the existence of two conserved charges, the usual charge and the dipole charge,
	\begin{equation}
		Q=\int_{space}\,\,j_0,\qquad Q^i=\int_{space}\,\,x^ij_0\label{k15}
	\end{equation}
	This is reminiscent of fractonic theories, the characteristic feature of which is the presence of higher derivatives. In the present case such terms are absent.\\
	It is tempting to ask whether a general formulation is possible to generate the currents (\ref{k11a},\ref{k11b}). Indeed a viable definition is given by,
	\begin{align}
		J_i&=w j_i-(\partial_k w)\,j_{ki}=\frac{\partial\mathcal{L}}{\partial(\partial_i\phi)}\delta\phi+\phi(\partial_i\,w)\label{k16a}\\
		J^0&=w\,j^0=\frac{\partial\mathcal{L}}{\partial(\partial_0\phi)}\delta\phi\label{k16b}
	\end{align}
	while there is no change in $J^0$ defined as in (\ref{k9a}), the vector current yields,
	\begin{equation}
		wj_i-\partial_kw\,j_{ki}=-(\partial_i\phi)\,w+\phi\,(\partial_iw)\label{k17}
	\end{equation}
	leading to,
	\begin{equation}
		j_i=-\partial_i\phi,\qquad j_{ki}=-\phi\delta_{ki}\label{k18}
	\end{equation}
	reproducing (\ref{k11b}) and (\ref{k13}).\\
	The equation for current conservation of $J_0,J_i$ obtained from the defining equation (\ref{k16a},\ref{k16b}) is given by,
	\begin{equation}
		\partial_0J_0-\partial_iJ_i=-\delta\mathcal{L}-\partial_i(\phi\partial_i\,w)=0\label{k19}
	\end{equation}
	that follows from (\ref{k8}). The standard vector current conservation is deduced by substituting the explicit values of $J_0,J_i$ in (\ref{k19}),
	\begin{equation}
		\partial_0(wj_0)-\partial_i(wj_i-(\partial_kw)\,j_{ki})=0\label{k20}
	\end{equation}
	
	Extracting $w$ out from the parenthesis, followed by a little algebra immediately yields (\ref{k12}).\\\\
	This derivation, however, conceals the fact that there are two conservation laws related to the charges $Q,Q^i$ (\ref{k15}). To reveal this, we open $w$ in favour of $c,c_i$ using (\ref{k6}). Then,
	\begin{equation}
		\partial_0[(c+c_jx_j)j_0]-\partial_i[(c+c_jx_j)j_i-\partial_k(c+c_jx_j)j_{ki}]=0\label{k21}
	\end{equation}
	Equating terms proportional to $c$ yields the standard conservation law (\ref{k12}) while equating those of $c_j$ yields,
	\begin{equation}
		\partial_0(\tilde{j}_{oj})-\partial_i(\tilde{j}_{ij}-j_{ij})=0\label{k22}
	\end{equation}
	where,
	\begin{equation}
		\tilde{j}_{oj}=x_j j_0,\qquad \tilde{j}_{ij}=x_j j_i\label{k23}
	\end{equation}
	and leads to the conservation of the dipole charge $Q^j$, defined in (\ref{k15}).
	\subsection{Nonrelativistic Maxwell theory}
	It is feasible to develop a similar algorithm for the nonrelativistic Maxwell field. The new point here is the interplay between shift and gauge symmetries. The dynamics is governed by the Lagrangian\footnote{$C_1$ and $C_2$ are, in general, different},
	\begin{equation}
		\mathcal{L}=\frac{C_1}{2}F_{oi}^2-\frac{C_2}{4}F_{ij}^2\label{k24}
	\end{equation}
	where
	\begin{equation}
		F_{0i}=\partial_0A_i-\partial_iA_0,\quad F_{ij}=\partial_iA_j-\partial_jA_i\label{k25}
	\end{equation}
	The equations of motion are given by,
	\begin{equation}
		C_1\partial_0F_{0i}-C_2\partial_jF_{ji}=0,\quad \partial_iF_{0i}=0\label{z}
	\end{equation}
	where the second one is the Gauss constraint.\\
	There is a gauge symmetry,
	\begin{equation}
		\delta A_0=\partial_0 w,\quad \delta A_i=\partial_iw\label{k26}
	\end{equation}
	that results in a conserved Noether current,
	\begin{align}
		j^0&=\frac{\partial\mathcal{L}}{\partial(\partial_0A_i)}\delta A_i=C_1F_{0i}(\partial_iw)\label{k27a}\\
		j_i&=\frac{\partial\mathcal{L}}{\partial(\partial_iA_j)}\delta  A_j+\frac{\partial\mathcal{L}}{\partial(\partial_iA_0)}\delta   A_0=-C_2F_{ij}(\partial_jw) -C_1F_{0i}(\partial_0w)\label{k27b}
	\end{align}
	which is easily verified,
	\begin{equation}
		\partial_0j_0-\partial_ij_i=(C_2\partial_iF_{ij}-C_1\partial_0F_{0j})(\partial_jw)+C_1\partial_iF_{0i}\partial_0w=0\label{k28}
	\end{equation}
	by using the equations of motion.\\
	Gauge symmetry, regarded as an origin of the masslessness of the photon, is translatable to the linear shift symmetry by choosing,
	\begin{equation}
		w=c_ix_i+\frac{1}{2}c_{ij}x_ix_j\label{k29}
	\end{equation}
	so that,
	\begin{equation}
		\delta A_i=c_i+c_{ij}x_j,\quad \delta A_0=0\label{k30}
	\end{equation}
	analogous to (\ref{k6}). Thus the gapless photon mode is interpreted as a consequence of gauge symmetry or shift symmetry similar to Goldstone bosons. Note that although, in general, $c_{ij}$ has both a symmetric part and an antisymmetric component, only the former contributes to \eqref{k29}. \\
	As we found for the scalar field there are other ways of defining currents. We adopt the same algorithm as before. It is useful to express $c_{ij}$ in terms of its symmetric $(s_{ij})$ and antisymmetric $(a_{ij})$ parts,
	\begin{equation}
		s_{ij}=\frac{1}{2}(c_{ij}+c_{ji}),\qquad a_{ij}=\frac{1}{2}(c_{ij}-c_{ji})\label{d1}
	\end{equation}
	which implies, 
	\begin{equation}
		\delta A_i=c_i+s_{ij}x_j+a_{ij}x_j=w_i(x)\label{d2}
	\end{equation}
	For the above shift the lagrangian (\ref{k24}) changes as,
	\begin{equation}
		\delta\mathcal{L}=2C_2\partial_i(a_{ij}A_j)=C_2a_{ij}F_{ij}\label{d3}
	\end{equation}
	and hence is quasi-invariant as well as gauge invariant.\\\\
 It is pertinent to observe that, retaining only the symmetric part $s_{ij}$ in \eqref{d3}, as dictated by the structure \eqref{k29}, would yield $\delta\mathcal{L}=0$ and there would be no boundary term. Exact gauge invariance follows which shows its direct correspondence with shift invariance. This is to be expected if one works with standard theories without higher derivative terms in the action.\\\\
 Since our motivation is to go beyond usual invariances by involving higher derivative terms, we envisage a wider possibility here itself by demanding a general form for $\delta A_i$ given by \eqref{d2} that includes both $s_{ij}$ and $a_{ij}$. The actual implementation of these ideas in the context of scalar field theories with higher derivatives is done in next (sub) section.\\\\
 Following our general algorithm the currents are defined analogous to (\ref{k16a},\ref{k16b}),
	\begin{align}
		J^0&=w_jj^{0j}=\frac{\partial\mathcal{L}}{\partial(\partial_0A_i)}\delta A_i=C_1F_{0i}w_i\label{d4a}\\
		J_i&=w_jj_{ij}-(\partial_kw_j)j_{ijk}=\frac{\partial\mathcal{L}}{\partial(\partial_iA_j)}\delta A_j-2C_2a_{ij}A_j\label{d4b}
	\end{align}
	where the boundary term (\ref{d3}) has been included in the expression for $J_i$. While (\ref{d4a}) yields,
	\begin{equation}
		j^{0j}=-j^j_0=C_1F_{0j}\label{y}
	\end{equation}
	simplifying (\ref{d4b}) leads to, 
	\begin{equation}
		w_jj_{ij}-(\partial_kw_j)j_{ijk}=-C_2F_{ij}w_j-2C_2a_{ij}A_j\label{d5}
	\end{equation}
	which yields,
	\begin{align}
		&j_{ij}=-C_2F_{ij}\label{d6a}\\
		&j_{ijk}=C_2(-\delta_{ik}A_j+\delta_{ij}A_k)\label{d6b}
	\end{align}
	From the above relations we get a condition,
	\begin{equation}
		\partial_ij_{ijk}=C_2F_{jk}=-j_{jk}\label{d7}
	\end{equation}
	analogous to (\ref{k13}).\\
	The conservation laws are given by,
	\begin{equation}
		\partial_0J_0-\partial_iJ_i=\partial_0(w_jj_{0j})-\partial_i\Big(w_jj_{ij}-(\partial_kw_j)j_{ijk}\Big)\label{d8}
	\end{equation}
	Abstracting out $w_j$ we find,
	\begin{equation}
		w_j(\partial_0j_{0j}-\partial_ij_{ij})=0\label{d9}
	\end{equation}
	on exploiting (\ref{d7}) and the linearity of $w_j$ (\ref{d2}). A direct way to verify (\ref{d9}) is to use (\ref{y}) and (\ref{d7}) which just yields the onshell condition (\ref{z}).\\
	As before, opening up $w_j$ yields another conservation law,
	\begin{equation}
		\partial_0[(c_j+s_{jk}x_k+a_{jk}x_k)j_{0j}]-\partial_i[(c_j+s_{jk}x_k+a_{jk}x_k)j_{ij}]+\partial_i[(s_{jk}+a_{jk})j_{ijk}]=0\label{d10}
	\end{equation}
	Equating coefficients of $c_j$ reproduces (\ref{d9}). Likewise, equating coefficients of $(s_{jk}+a_{jk})$ yields,
	\begin{equation}
		\partial_0(x_kj_{0j})-\partial_i[x_kj_{ij}-j_{ijk}]=0\label{d11}
	\end{equation}
	and leads to the conservation of a dipole like (vector) charge,
	\begin{equation}
		Q_{kj}=\int_{space}\,\, x_kj_{0j}\label{d12}
	\end{equation}
	The field momentum and dipole charge satisfy a generalised Heisenberg algebra,
	\begin{equation}
		[P_i,Q_{kj}]=\Big[\int\,dx\,\pi_l\partial_iA_l(x),\int\,dy\,y_kj_{0j}(y)\Big]=-C_1\Big[\int\,dx\,dy\pi_l\partial_i A_l(x),y_kF_{0j}(y)\Big]\label{d13}
	\end{equation}
	Recalling that $F_{0j}$ is the momentum canonically conjugate to $A_j$,
	\begin{equation}
		\pi_j=\frac{\partial\mathcal{L}}{\partial\dot{A}_j}=C_1F_{0j}\label{d14}
	\end{equation}
	we obtain,
	\begin{equation}
		[P_i,Q_{kj}]=\delta_{ik}Q_j\label{d15}    
	\end{equation}
	where $Q_j$ is the (vector) charge,
	\begin{equation}
		Q_j=\int_{space}\,\,j_{0j}\label{d16}
	\end{equation}
	Appearance of higher rank tensor currents in standard Maxwell's theory is a precursor to similar currents in higher derivative scalar field theories, both real and complex. Since it forms a central part of our analysis, we discuss this in some details.
	\subsection{Higher derivative scalar field theories}
 The simplest higher derivative extension of the usual theory (\ref{k7}) is given by,
	\begin{equation}
		\mathcal{L}=\frac{1}{2}\dot{\phi}^2-\frac{1}{2}(\partial_i\partial_j\phi)(\partial_i\partial_j\phi) \label{d17}
	\end{equation}
	which is a typical example of a Lifshitz type model with a scaling symmetry,
	\begin{equation}
		\delta x^i = \lambda x^i \,; \,\,\, \delta t = z \lambda t\label{d18} \,\,\, (z=\lambda)
	\end{equation}
	where $z$  characterises the degree of anisotropy between space and time.
	
	A cubic shift in the coordinates,
	\begin{equation}
		\phi \,\to\, \phi+w,\quad w=c+c_ix_i+\frac{1}{2}c_{ij}x_ix_j+\frac{1}{3}c_{ijk}x_ix_jx_k\label{d19}
	\end{equation}
	leaves the lagrangian quasi-invariant, since,
	\begin{equation}
		\delta \mathcal{L}=-\partial_i[\partial_j\phi\partial_i\partial_jw]+\partial_j[\phi\partial^2\partial_jw]-\phi(\partial^2)^2 w\label{d20}
	\end{equation}
	where the last term drops out since $w= O (x^3)$.\\
	Since the lagrangian involves double derivatives, the Noether currents are defined as,
	\begin{align}
		J^0&=\frac{\partial\mathcal{L}}{\partial(\partial_0\phi)}\delta\phi=\dot{\phi}w\nonumber\\
		J^i&=J_i=\frac{\partial\mathcal{L}}{\partial(\partial_i\phi)}\delta\phi+\frac{\partial\mathcal{L}}{\partial(\partial_i\partial_j\phi)}\partial_j(\delta\phi)-\Big(\partial_j\frac{\partial\mathcal{L}}{\partial(\partial_i\partial_j\phi)}\Big)\delta\phi+\partial_j\phi\partial_i\partial_jw-\phi\partial^2\partial_iw\nonumber\\
		&=-(\partial_i\partial_j\phi)(\partial_jw)+\partial_j(\partial_i\partial_j\phi)w+(\partial_j\phi)(\partial_i\partial_j w)-\phi(\partial^2\partial_i w)\label{d21}
	\end{align}
	where the boundary terms have been inserted. Current conservation now follows,
	\begin{equation}
		\partial_0J_0-\partial_iJ_i=-w(\Ddot{\phi}+(\partial^2)^2\phi)+\phi(\partial^2)^2w\label{d22}
	\end{equation}
	which vanishes on-shell.\\ \\
	The other (higher order) currents and their associated conservation laws follow on using our definition,
	\begin{equation}
		J^0= wj^0=\frac{\partial\mathcal{L}}{\partial(\partial_0\phi)}\delta\phi=\dot{\phi}w\label{d23}
	\end{equation}
	which yields,
	\begin{equation}
		j^0=\dot{\phi}\label{d24}
	\end{equation}
	The space current is given by,
	\begin{align}
		J_i&=wj_i-(\partial_j w)j_{ji}+(\partial_j\partial_k w)j_{jki}-(\partial_j\partial_k\partial_l w)j_{jkli}\nonumber\\
		&=\frac{\partial\mathcal{L}}{\partial(\partial_i\partial_j\phi)}\partial_j(\delta\phi)-\Big(\partial_j\frac{\partial\mathcal{L}}{\partial(\partial_i\partial_j\phi)}\Big)\delta\phi+(\partial_j\phi)\partial_i\partial_jw-\phi\partial^2\partial_i w\nonumber\\
		&=-(\partial_i\partial_j\phi)\partial_j w+\partial_j(\partial_i\partial_j\phi)w+(\partial_j\phi)\partial_i\partial_j w-\phi \partial^2\partial_i w\label{d25}
	\end{align}
	Comparing coefficients of $w$ and its derivatives yields,
	\begin{align}
		j_i&=\partial_i\partial^2\phi\label{d26a}\\
		j_{ji}&=\partial_i\partial_j\phi=j_{ij}\label{d26b}\\
		j_{jki}&=\frac{1}{2}(\delta_{ki}\partial_j\phi+\delta_{ij}\partial_k\phi)=j_{kji}\label{d26c}\\
		j_{jkli}&=\frac{1}{3}(\delta_{jk}\delta_{li}+\delta_{jl}\delta_{ki}+\delta_{kl}\delta_{ji})\phi\label{d26d}
	\end{align}
	The following set of  identities are obtained from the above relations,
	\begin{equation}
		j_i=\partial_jj_{ij}=\partial_j\partial_kj_{ijk}=\partial_j\partial_k\partial_lj_{ijkl}\label{d27}
	\end{equation}
	Besides this chain, there exists another identity,
	\begin{equation}
		j_{jk}=\partial_ij_{jki}\label{m1}
	\end{equation}
	The above identities are useful for proving conservation laws, as shown below.
	It is important to point out that the first relation in (\ref{d27}), which connects a vector current to a derivative of a rank two current, and the identity (\ref{m1}) that connects a rank two current to a derivative of rank three current contain the genesis of the duality among different gauge theories, as demonstrated later. Current conservation is now verified as,
	\begin{equation}
		\partial_0J_0-\partial_iJ_i=\partial_0(wj_0)-\partial_iJ_i=\partial_0(-\dot{\phi}w)-\partial_i\Big[wj_i-(\partial_jw)j_{ji}+(\partial_j\partial_kw)j_{jki}-(\partial_j\partial_k\partial_lw)j_{jkli}\Big]=0\label{d28}
	\end{equation}
	Factoring out $w$ and exploiting the identities (\ref{d27}) yields, 
	\begin{equation}
		\partial_0J_0-\partial_iJ_i=-w(\Ddot{\phi}+(\partial^2)^2\phi)+(\partial_i\partial_j\partial_k\partial_l w)j_{jkli}\label{d29}    
	\end{equation}
	which vanishes on shell and the fact that $w$ is cubic in the coordinates.\\
	This derivation hides the whole tower of conservation laws that may be revealed by opening up $w$ using (\ref{d19}). Then equating one by one the factors multiplying parameters $c,c_i$ etc, we have, from (\ref{d28}),
	\begin{align}
		\partial_0j_0-\partial_ij_i=&0\nonumber\\
		\partial_0(x_ij_0)-\partial_j(x_ij_j-j_{ij})=0\label{d30}
	\end{align}
	While the first leads to the conservation of the usual charge $Q=\int_{space} j_0$, the second yields the conservation of the dipole charge,
	\begin{equation}
		Q_i=\int_{space} \,\,x_ij_0\label{d31}
	\end{equation}
	Similarly, equating coefficients of $c_{ij}$ yields,
	\begin{equation}
		\partial_0 \Big(\frac{1}{2}x_ix_jj_0\Big)-\partial_k\Big[\frac{1}{2}x_ix_jj_k-\frac{1}{2}(x_ij_{jk}+x_jj_{ik})+j_{ijk}\Big]=0\label{d32}
	\end{equation}
	which leads to the conservation of the quadrupole charge,
	\begin{equation}
		Q_{ij}= \int_{space} x_ix_jj_0\label{d33}
	\end{equation}
	A direct check of (\ref{d32}) can be done by first simplifying the left side 
	\begin{equation}
		\frac{1}{2}x_ix_j(\partial_0j_0-\partial_kj_k)-\frac{1}{2}(x_jj_i+x_ij_j)+\frac{1}{2}(x_i\partial_kj_{jk}+x_j\partial_kj_{ik})+\frac{1}{2}(j_{ji}+j_{ij})-\partial_kj_{ijk}=0\label{d34}
	\end{equation}
	followed by exploiting the identities (\ref{d26a},\ref{d27}). The various terms cancel out while the first bracket vanishes by usual vector current conservation.\\
	Finally, equating coefficients of $c_{ijk}$ in (\ref{d28}) yields the conservation law,
	\begin{equation}
		\partial_0(x_ix_jx_kj_0)-\partial_l\Big[x_ix_jx_kj_l-\partial_m(x_ix_jx_k)j_{ml}+\partial_m\partial_n(x_ix_jx_k)j_{mnl}-\partial_m\partial_n\partial_p(x_ix_jx_k)j_{mnpl}\Big]=0\label{d35}
	\end{equation}
	The octupole charge is now conserved, 
	\begin{equation}
		Q_{ijk}=\int_{space}\, x_ix_jx_kj_0,\quad \dot{Q}_{ijk}=0\label{d36}
	\end{equation}
	The various multipole charges form a closed algebra with the the symmetry generators which will be discussed now.
	\subsection{The multipole algebra}
	The symmetry algebra of charges is obtained here for different cases like dipole, quadrupole etc.\\
	Let us first begin with the dipole charge $Q^i$. 	The field momentum (translation generator) $P_i$ and dipole charge $Q^j$ satisfy the Lie algebra of the Heisenberg group. To see this we compute,
	\begin{equation}
		[P_i,Q^j]=\Big[\int\,dx\, \pi\partial_i\phi(x),\int\,dy\,y^jj_0(y)\big]
	\end{equation}
	Recalling that $j^0$ (\ref{k11a}) is the canonical momenta $\pi$ and using the canonical bracket among $\phi-\pi$, we obtain, 
	\begin{equation}
		[P_i,Q^j]=-i\delta_i\,^j\,Q\label{cc2}
	\end{equation}which yields the cherished form.\\
	Similarly the angular momentum-dipole charge satisfies the algebra,
	\begin{equation}
	    [M_{ij},Q_k]=\Big[\int\,\,dx(x_i\pi\partial_j\phi-x_j\pi\partial_i\phi),\int\,\,dy\,y^kj_0(y)\Big]=-i(\delta_{jk}Q_i-\delta_{ik}Q_j)\label{cc1}
	\end{equation}
	The monopole charge $Q$ in (\ref{cc2}) and the dipole charge in (\ref{cc1}) appear as central extensions of the algebra. Similar results follow for higher pole charges. Specifically, for the quadrupole charge,
	\begin{align}
	    [P_i,Q_{kj}]&=-i(\delta_{ik}Q_j+\delta_{ij}Q_k)\label{B+}\\
	    [M_{ij},Q_{kl}]&=-i(\delta_{ik}Q_{jl}+\delta_{il}Q_{jk}-\delta_{jk}Q_{il}-\delta_{jl}Q_{ik})
	\end{align}
	and for the octupole charge,
	\begin{align}
	    [P_i,Q_{jkl}]&=-i(\delta_{ik}Q_{jl}+\delta_{ij}Q_{kl}+\delta_{il}Q_{jk})\label{B-}\\
	    [M_{ij},Q_{klm}]&=-i(\delta_{jk}Q_{ilm}+\delta_{jl}Q_{ikm}+\delta_{jm}Q_{ikl}-\delta_{ik}Q_{jlm}-\delta_{il}Q_{jkm}-\delta_{im}Q_{jkl})
	\end{align}
	The entire algebra is similar to the Heisenberg algebra and its extensions like, for instance,
	\begin{align}
	    [p_i,q_j]&=-i\delta_{ij}\nonumber\\
	    [p_i,q_jq_k]&=-i(\delta_{ij}q_k+\delta_{ik}q_j)\nonumber\\
	    [p_i,q_jq_kq_l]&=-i(\delta_{ij}q_kq_l+\delta_{ik}q_jq_l+\delta_{il}q_jq_k)
	\end{align}
	which are manifested in (\ref{cc2}), (\ref{B+}) and (\ref{B-}).\\
	Likewise the angular momentum algebra is another such manifestation,
	\begin{equation}
	    [q_ip_j-q_jp_i,q_k]=-i(\delta_{jk}q_i-\delta_{ik}q_j)
	\end{equation}
	whose analougue can be found in (\ref{cc1}) and which can be extended to the higher pole charges.
	\subsection{Alternative forms for currents}
	As is known, expressions for currents are not unique and may differ by local counterterms. Indeed this happens here also if we consider an equivalent form of the lagrangian (\ref{d17}),
	\begin{equation}
		\mathcal{L}_{eq}=\frac{1}{2}\dot{\phi}^2-\frac{1}{2}(\partial^2\phi)^2\label{m2}
	\end{equation}
	Under the cubic shift (\ref{d19}), the lagrangian changes by a total derivative, 
	\begin{equation}
		\delta\mathcal{L}_{eq}=\partial_i[\phi\partial_i\partial^2w-\partial_i\phi\partial^2w]\label{m3}
	\end{equation}
	The Noether currents (\ref{d21}) are now defined as,
	\begin{align}
		\tilde{J}^0&=\frac{\partial\mathcal{L}_{eq}}{\partial(\partial_0\phi)}\delta\phi=\dot{\phi}w\nonumber\\
		\tilde{J}^i&=\tilde{J}_i=\frac{\partial\mathcal{L}_{eq}}{\partial(\partial_i\phi)}\delta\phi+\frac{\partial\mathcal{L}_{eq}}{\partial(\partial_i\partial_j\phi)}\partial_j(\delta\phi)-\Big(\partial_j\frac{\partial\mathcal{L}_{eq}}{\partial(\partial_i\partial_j\phi)}\Big)\delta\phi-\phi\partial_i\partial^2\omega+\partial_i\phi\partial^2w\nonumber\\
		&=\partial_i\partial^2\phi-(\partial^2\phi)\partial_iw+(\partial_i\phi)\partial^2w-\phi(\partial_i\partial^2w)\label{m4}
	\end{align}
	While $J^0$ in (\ref{d21}) and $\tilde{J}^0$ are identical, the spatial currents differ by local counterterms,
	\begin{equation}
		\tilde{J}^i-J^i=\partial_j\Lambda_{ji}
	\end{equation}
	where,
	\begin{equation}
		\Lambda_{ji}=-\Lambda_{ij}=\partial_jw\partial_i\phi-\partial_iw\partial_j\phi\label{m5}
	\end{equation}
	which ensures conservation of $\tilde{J}$ current since,
	\begin{equation}
		\partial_i\tilde{J}^i=\partial_iJ^i\label{m6}
	\end{equation}
	The tensor currents, analogous to (\ref{d25}), are now obtained by adopting the definition (\ref{m4}), so that,
	\begin{align}
		&w\tilde{j}_i-\partial_jw\tilde{j}_{ji}+\partial_j\partial_kw\tilde{j}_{jki}-\partial_j\partial_k\partial_lwj_{jkli}\nonumber\\
		&=(\partial_i\partial^2\phi) w-(\partial^2\phi)\partial_iw+(\partial_i\phi)\partial^2w-\phi(\partial_i\partial^2w)\label{m7}
	\end{align}
	This yields,
	\begin{align}
		\tilde{j}_i&=\partial_i\partial^2\phi\nonumber\\
		\tilde{j}_{ji}&=\delta_{ji}\partial^2\phi=\tilde{j}_{ij}\nonumber\\
		\tilde{j}_{jki}&=\delta_{kj}\partial_i\phi=\tilde{j}_{kji}\nonumber\\
		\tilde{j}_{jkli}&=\frac{1}{3}(\delta_{jk}\delta_{li}+\delta_{jl}\delta_{ki}+\delta_{kl}\delta_{ji})\phi \label{m8}
	\end{align}
	While $j_i$ and $\tilde{j}_i$, $j_{jkil}$ and $\tilde{j}_{jkil}$ exactly match, the other two tensor currents differ in a way such that conservation is retained.
	\begin{align}
		& \partial_jj_{ji}=\partial_j\tilde{j}_{ji}=\partial_i\partial^2\phi\nonumber\\
		&\partial_j\partial_kj_{jki}=\partial_j\partial_k\tilde{j}_{jki}=\partial_i\partial^2\phi\label{m9}
	\end{align}
	Finally, the identities (\ref{d27}) and (\ref{m1}) are also preserved,
	\begin{align}
		\tilde{j}_i&=\partial_j\tilde{j}_{ji}=\partial_j\partial_k\tilde{j}_{ijk}=\partial_j\partial_k\partial_l\tilde{j}_{ijkl}\nonumber\\
		\tilde{j}_{jk}&=\partial_i\tilde{j}_{jki}\label{m10}
	\end{align}
	which is necessary to establish current conservation.\\
	It may be mentioned that the structure (\ref{d26a}-\ref{d26d}) or (\ref{m8}) is generic while the identities (\ref{d27}) and (\ref{m1}) are exactly retained for any higher derivative theory that is a generalisation of (\ref{d17}) or (\ref{m2}). For instance if we consider the lagrangian,
	\begin{equation}
		\mathcal{L}=\frac{1}{2}\dot{\phi}^2-\frac{1}{2}(\partial_i\partial_j\partial_k\phi)(\partial_i\partial_j\partial_k\phi)\label{m12}
	\end{equation}
	then, under a quintic shift of coordinates,
	\begin{equation}
		\phi\to\phi+w,\,\,w=c+c_ix_i+\frac{1}{2}c_{ij}x_ix_j+\frac{1}{3}c_{ijk}x_ix_jx_k+\frac{1}{4}c_{ijkl}x_ix_jx_kx_l+\frac{1}{5}c_{ijklm}x_ix_jx_kx_lx_m\label{m13}
	\end{equation}
	it is left quasi invariant,
	\begin{equation}
		\delta\mathcal{L}=\partial_i\Lambda_i-\phi(\partial^2)^3w\label{m14}
	\end{equation}
	where
	\begin{equation*}
		\Lambda_i=(\partial_j\partial_k\phi)(\partial_i\partial_j\partial_kw)-(\partial_k\phi)(\partial^2\partial_i\partial_k w)+\phi(\partial^2)^2\partial_i w
	\end{equation*}
	Following the algorithm developed here the basic conservation law is expressed as,
	\begin{equation}
		\partial_0J_0-\partial_iJ_i=\partial_0(wj_0)-\partial_iJ_i=0\label{m15}
	\end{equation}
	where
	\begin{equation}
		J_i=wj_i-(\partial_jw)j_{ji}+(\partial_j\partial_kw) j_{jki}-(\partial_j\partial_k\partial_l w)j_{jkli}+(\partial_j\partial_k\partial_l \partial_mw)j_{jklmi}-(\partial_j\partial_k\partial_l \partial_m\partial_n w) j_{jklmni}\label{m16}
	\end{equation}
	and the individual expressions are given by,
	\begin{align}
		j_i=&-\partial_i\partial^4\phi,\quad j_{ji}=-\partial_i\partial_j\partial^2\phi,\quad j_{jki}=-\partial_i \partial_j\partial_k\phi\nonumber\\
		j_{jkli}=&-\frac{1}{3}(\delta_{il}\partial_j\partial_k\phi+\delta_{ik}\partial_j\partial_l\phi+\delta_{ij}\partial_k\partial_l\phi)\nonumber\\
		j_{jklmi}=&....\label{m17}
	\end{align}
	The tower of conservation laws corresponding to dipole and higher pole symmetries can be calculated as shown earlier and are based on identities similar to (\ref{d27}).
	\begin{equation}
		j_i=\partial_jj_{ji}=\partial_j\partial_kj_{jki}=\partial_j\partial_k\partial_lj_{jkli}=\partial_j\partial_k\partial_l\partial_mj_{jklmi}=\partial_j\partial_k\partial_l\partial_m\partial_nj_{jklmni}\label{m18}
	\end{equation}
	Also, there exist intermediate identities like,
	\begin{equation}
		j_{jk}=\partial_ij_{ijk}
	\end{equation}
	as happened for the theory defined by (\ref{d17}).
	\section{From shift symmetries to gauge symmetries}
	The analysis for polynomial shift symmetries in scalar field theories done in the previous section is now generalised to complex scalars. This will allow interactions to be introduced by coupling with gauge fields-either vector or higher rank tensors. Such theories have recently been profusely discussed in the literature.\\
	A complex scalar $\Phi$ can be polar decomposed as,
	\begin{equation*}
		\Phi=\rho e^{i\phi}    
	\end{equation*}
	where $\rho, \phi$ are two real scalars. Then, under the shift,
	\begin{equation}
		\phi \to \phi +w\label{AA}
	\end{equation}
	where $w$ is an arbitrary space dependent polynomial,
	\begin{equation}
		\Phi \to e^{iw}\Phi\label{X}
	\end{equation}
	which is the usual abelian gauge transformation on the complex scalar $\Phi$. Then,
	\begin{equation}
		ln \Phi\,\to\, ln \Phi+i w
	\end{equation}
	Comparison with (\ref{AA}) shows that, as far as transformations are concerned, there exists the map,
	\begin{equation}
		i\phi \to ln \Phi\label{BB}
	\end{equation}
	With this map, we can construct suitable actions of complex scalars begining from real scalars such that the symmetries are preserved. This is illustrated by examples.\\\\
	As a first case we construct an action for complex scalars that is exactly invariant under a linear shift like (\ref{k6}). From (\ref{d17}) and (\ref{d20}) it is clear that (\ref{d17}) is exactly invariant ($\delta\mathcal{L}=0$) under a linear shift. Thus our cherished goal will be achieved by starting from (\ref{d17}) and using (\ref{BB}) to get the corresponding theory. We find
	\begin{equation}
		i\partial_i\partial_j\phi\,\,\to\,\,\frac{\Phi\partial_i\partial_j\Phi-\partial_i\Phi\partial_j\Phi}{\Phi^2}
	\end{equation}
	and hence,
	\begin{equation}
	\mathcal{L}=(\partial_i\partial_j\phi)^2=|i\partial_i\partial_j\phi|^2\,\to\,\frac{|\Phi\partial_i\partial_j\Phi-\partial_i\Phi\partial_j\Phi|^2}{|\Phi^2|^2}\label{r1}
	\end{equation}
	This implies that the expression on the right will be invariant under the gauge transformation (\ref{X}) since the corresponding expression on the left side is invariant under the linear shift (\ref{AA}). In either case $w$ is given by (\ref{k6}).	Since $|\Phi^2|^2$ is invariant under a general U(1) gauge transformation, we just take the numerator in (\ref{r1}) as the desired lagrangian. This reproduces the expression for the lagrangian invariant under a linearly shifted U(1) gauge transformation, as found in the literature.\\	The conserved currents may likewise be obtained by a direct application of the equality connecting the $\phi$ and $\Phi$ fields,
	\begin{equation}
		\phi=-\frac{i}{2}ln\Phi+c.c\label{nn7}
	\end{equation}
	Then it follows from (\ref{d26b}),
	\begin{equation}
		j_{ji}=j_{ij}=\partial_i\partial_j\phi=\frac{-i[{\Phi^*}^2(\Phi\partial_i\partial_j\Phi-\partial_i\Phi\partial_j\Phi)-\Phi^2(\Phi^*\partial_i\partial_j\Phi^*-\partial_i\Phi^*\partial_j\Phi^*)]}{2|\Phi|^4}\label{105}
	\end{equation}
	while the vector current follows by simply using,
	\begin{equation}
		j_i=\partial_jj_{ji}\label{W}
	\end{equation}
	Now we shall similarly construct an action for complex scalar theory which is invariant under a quadratic shift. From (\ref{m12}) and (\ref{m14}) it can be understood that (\ref{m12}) is invariant under a quadratic shift. So we shall now use the map (\ref{BB}) in (\ref{m12}) to achieve the corresponding complex scalar theory.
	\begin{equation}
		i\partial_i\partial_j\partial_k\phi\, \to\, \frac{\Phi^2\partial_i\partial_j\partial_k\Phi+2\partial_i\Phi\partial_j\Phi\partial_k\Phi-\Phi(\partial_i\Phi\partial_j\partial_k\Phi+\partial_j\Phi\partial_k\partial_i\Phi+\partial_k\Phi\partial_i\partial_j\Phi)}{\Phi^3}\label{m21}
	\end{equation}
	and hence,
	\begin{equation}
	\mathcal{L}=(\partial_i\partial_j\partial_k\phi)^2=|i\partial_i\partial_j\partial_k\phi|^2\,\to\, \frac{\Big|\Phi^2\partial_i\partial_j\partial_k\Phi+2\partial_i\Phi\partial_j\Phi\partial_k\Phi-\Phi(\partial_i\Phi\partial_j\partial_k\Phi+\partial_j\Phi\partial_k\partial_i\Phi+\partial_k\Phi\partial_i\partial_j\Phi)\Big|^2}{|\Phi|^6}\label{m22}
	\end{equation}
 	As before, one may drop the $|\Phi|^6$ in the denominator, retaining only the numerator. It reproduces the well known expression for the lagrangian invariant under a quadratically shifted U(1) gauge transformation. \\
The conserved current can now be derived from (\ref{m17}) using the map between $\phi$ and $\Phi$ (\ref{nn7}) as
\begin{equation}
    j_{ijk}=\partial_i\partial_j\partial_k\phi=\frac{-i{\Phi^*}^3\Big[\Phi^2\partial_i\partial_j\partial_k\Phi+2\partial_i\Phi\partial_j\Phi\partial_k\Phi-\Phi(\partial_i\Phi\partial_j\partial_k\Phi+\partial_j\Phi\partial_k\partial_i\Phi+\partial_k\Phi\partial_i\partial_j\Phi)\Big]+c.c}{2|\Phi|^6}
\end{equation}
while the lower rank currents follow by simplifying
\begin{equation*}
    j_i=\partial_j\partial_kj_{ijk},\quad j_{ij}=\partial_kj_{ijk}
\end{equation*}
Here we derived the lagrangians and corresponding tensor currents for the matter sector in higher derivative complex scalar theories invariant under a gauge transformation, where the gauge parameter is a space dependent linear or quadratic shift. This was achieved by exploiting a map from real to complex scalar field using the structure of lagrangians and currents derived for real scalar fields in the previous section. We shall use these lagrangians in our upcoming sections to construct a full interacting gauge theory.
	\section{Dual description of theory with space dependent linear shift symmetry}
	Here we briefly review the coupling prescription for a lagrangian involving complex scalar fields invariant under space dependent linear shift. It is also shown that, interactions can be introduced either using usual gauge (vector) fields or, alternatively, using second rank symmetric tensor gauge fields. In the literature \cite{pretko2020,hoyos,seiberg} only the second possibility has been discussed at length.\\
	The gauge transformation on a scalar field is given by
	\begin{equation}
		\Phi(x)\,\,\to\,\,\Phi'(x)=e^{iw(x)}\Phi(x),\quad \Phi^*(x)\,\,\to\,\,\Phi'^*(x)=e^{-iw(x)}\Phi^*(x)\label{a1}
	\end{equation} 
	where the transformation parameter $w$ depends linearly on the space coordinate $x_i$ (\ref{k6})\footnote{Rotational invariance is assumed.}. It has been shown that \cite{seiberg} the following scalar field lagrangian is invariant under the above transformation (\ref{a1});
	\begin{equation}
		\mathcal{L}=\frac{i}{2}\Phi^*\overleftrightarrow{\partial_0}\Phi- s\Big(\partial_i(\Phi^*\Phi)\Big)^2+it({\Phi^*}^2\partial_i\Phi\partial_i\Phi-\Phi^2\partial_i\Phi^*\partial_i\Phi^*)+u|\Phi\partial_i\partial_j\Phi-\partial_i\Phi\partial_j\Phi|^2-\lambda|\Phi|^4\label{r2}
	\end{equation}
	where $s,t$ and $u$ are some normalization factors. The first two terms and the last term in the above lagrangian are trivially invariant under (\ref{a1}), even when $w(x)$ is an arbitrary local parameter. Let us therefore concentrate on the $u$ term. This particular term was also obtained here in (\ref{r1}), apart from the $|\Phi^2|^2$ term which, incidentally, is trivially invariant under an arbitrary local transformation. The term with coefficient $u$ changes under a general transformation as
	\begin{equation}
		|\Phi\partial_i\partial_j\Phi-\partial_i\Phi\partial_j\Phi|^2 \to |\Phi\partial_i\partial_j\Phi-\partial_i\Phi\partial_j\Phi+i\Phi^2\partial_i\partial_jw|^2\label{r4}
	\end{equation}
	Invariance is achieved provided $w$ is linearly dependent on $x_i$, as given in (\ref{k6}). This is the reason behind taking the particular form of the lagrangian (\ref{r2}), which is invariant under (\ref{a1}) for $w$ having the form (\ref{k6}).\\
 The discussion so far has been confined to taking a specific form for $w(x)$, appearing in \eqref{a1}, that is either constant or linear in $x$. In either case, \eqref{r2} is invariant under the global transformation \eqref{a1} without the need for introducing any gauge field. We now wish to lift this global symmetry to a local symmetry. Thus let us take a general space-dependent parameter, which will obviously not keep the lagrangian invariant. So we shall have to add a compensating term in the lagrangian, preferably a gauge field using some prescription, which will render the invariance of the lagrangian. In this way gauge couplings can be introduced. As we shall show, this may be achieved using tensor gauge fields or vector gauge fields \cite{dual}. The complete theory comprising of the matter and gauge  (given by either vector or tensor) fields is now invariant under an arbitrary space-dependent transformation. The issue of global symmetries, for either matter or the gauge fields now becomes inconsequential.\\\\
	Using (\ref{r4}) it is easy to find a covariantization prescription involving a tensorial gauge field, for this part of the lagrangian. It is given by
	\begin{equation}
		\Phi\partial_i\partial_j\Phi-\partial_i\Phi\partial_j\Phi\quad\to \quad \Phi\partial_i\partial_j\Phi-\partial_i\Phi\partial_j\Phi-iA_{ij}\Phi^2\label{a3}
	\end{equation}
	so that the covariance can be achieved, under (\ref{a1}) and 
	\begin{equation}
		A_{ij}\to A_{ij}+\partial_i\partial_j w\label{a2}
	\end{equation}
	where $A_{ij}$ is a second rank symmetric tensor field.	It is seen that (\ref{a3}) transforms under (\ref{a1}) and (\ref{a2}) covariantly as
	\begin{equation}
		\Phi\partial_i\partial_j\Phi-\partial_i\Phi\partial_j\Phi-iA_{ij}\Phi^2\quad\to\quad e^{2iw(x)} (\Phi\partial_i\partial_j\Phi-\partial_i\Phi\partial_j\Phi-iA_{ij}\Phi^2)\label{a4}
	\end{equation}
	where the non covariant piece $\partial_i\partial_jw$ (see (\ref{r4})) is cancelled by a contribution from the transformation of $A_{ij}$. So the total lagrangian is given by
	\begin{equation}
		\mathcal{L}=\Big|\Phi\partial_i\partial_j\Phi-\partial_i\Phi\partial_j\Phi-iA_{ij}\Phi^2\Big|^2\label{h52}
	\end{equation}
	At this juncture we make an important observation. It may be shown that the non-standard gauging prescription (\ref{a3}) can be mapped to the standard minimal prescription using the usual vector gauge field $A_i$. The origin of this lies in the presence of conservation laws other than  usual charge conservation. It was shown in the previous section that, for linear  shift symmetry, there are two currents $j^i$ and $j^{ij}$ related by (\ref{k13})
	This kind of relation enables one to question whether the vector and the tensor fields are also related. Infact coupling the vector current with usual vector gauge field gives,
	\begin{equation}
		A_ij^i=A_i\partial_jj^{ij}=-\frac{1}{2}(\partial_jA_i+\partial_iA_j)j^{ij}=A_{ij}j^{ij}\label{B2}
	\end{equation}
	where we have dropped a boundary term. So the above relation enables us to identify immediately 
	\begin{equation}
		A_{ij}:=-\frac{1}{2}(\partial_iA_j+\partial_jA_i).\label{a22}
	\end{equation} 
	This identification is also compatible with transformations (\ref{a2}) and $\delta A_i=-\partial_i w$ \footnote{ Note the difference in sign from (\ref{k26}).}. 
	After obtaining the relation between a tensor and a vector gauge field we shall now discuss how one can use the minimal coupling prescription with $A_i$ instead of the non-standard prescription shown in (\ref{a3}) and can give the same interacting theory for linear shift symmetric lagrangian.\\
	For that we first replace the ordinary derivatives $ \partial_i$ with the covariant derivative $\partial_i+iA_i$, where $A_i$ is the usual vector field transforming as (\ref{k26})(see footnote 5),
\begin{equation}	(\partial_i+iA_i)\Phi\,\,\to\,\, e^{iw}(\partial_i+iA_i)\Phi	\end{equation}
	Now one has to symmetrize the double derivative term of the lagrangian, followed by covariantization of the derivative, demonstrated  below:
	\begin{align}
		&\Phi\partial_i\partial_j\Phi-\partial_i\Phi\partial_j\Phi\,\,\to\,\, \frac{1}{2}(\Phi\partial_i\partial_j\Phi+\Phi\partial_j\partial_i\Phi)-\partial_i\Phi\partial_j\Phi\nonumber\\
		&\to\,\, \frac{1}{2}\Big[\Phi(\partial_i+iA_i)(\partial_j+iA_j)\Phi+\Phi(\partial_j+iA_j)(\partial_i+iA_i)\Phi\Big]-(\partial_i+iA_i)\Phi(\partial_j+iA_j)\Phi\label{a5}
	\end{align}    
	On simplification the above expression yields
	\begin{equation}
		\Phi\partial_i\partial_j\Phi-\partial_i\Phi\partial_j\Phi\,\,\to\,\,\Phi\partial_i\partial_j\Phi-\partial_i\Phi\partial_j\Phi+\frac{i}{2}(\partial_iA_j+\partial_jA_i)\Phi^2\label{a6}	
	\end{equation}
	This expression exactly matches with the one in (\ref{a3}), provided we identify $A_{ij}$ as gien in (\ref{a22}) and the total lagrangian is now given by
	\begin{equation}
		\mathcal{L}=\Big|\Phi\partial_i\partial_j\Phi-\partial_i\Phi\partial_j\Phi+\frac{i}{2}(\partial_iA_j+\partial_jA_i)\Phi^2\Big|^2  \label{h51}
	\end{equation}
	So (\ref{a5}) gives a minimal prescription using usual vector field that gives the same result as that of (\ref{a3}), which is obtained in terms of a rank-2 symmetric tensor.
	The preceding analysis revealed the intriguing possibility of working with either vector or symmetric higher rank tensor gauge fields when gauging higher derivative complex scalar theories.
	\subsection{Gauge coupling from iterative Noether prescription}
	Here we first obtain the currents directly, adopting a definition analogous to (\ref{d25}). Then, following the iterative prescription \cite{deser} we derive the gauge couplings. The dual description, expressed in terms of a vector field or a tensor field, is reproduced. Let us consider the $u$ term in (\ref{r2}). The conserved currents are defined analogous to (\ref{d25}), recalling that the shift symmetry is only linear (\ref{k6}) \cite{noether}
	\begin{align}
	J^0 &=j^0=\frac{\partial\mathcal{L}}{\partial(\partial_0\Phi)}\delta\Phi+c.c=-\Phi^*\Phi\nonumber\\
		J^i&= wj^i-(\partial_jw)j^{ji}=\frac{\partial\mathcal{L}_0}{\partial(\partial_i\Phi)}\delta\Phi+\frac{\partial\mathcal{L}_0}{\partial(\partial_i\partial_j\Phi)}\partial_j(\delta\Phi)-\Big(\partial_j\frac{\partial\mathcal{L}_0}{\partial(\partial_i\partial_j\Phi)}\Big)\delta\Phi+c.c\label{h1}
	\end{align}
	Equating coefficients of $c,c_i$, we find,
	\begin{equation}
		j^i=\partial_jj_{ij}=i\partial_j\Big(\Phi^2(\Phi^*\partial_i\partial_j\Phi^*-\partial_i\Phi^*\partial_j\Phi^*)-{\Phi^*}^2(\Phi\partial_i\partial_j\Phi-\partial_i\Phi\partial_j\Phi)\Big)\label{h2}
	\end{equation}
	Modulo the multiplication factor $|\Phi|^4$, this current was earlier introduced in (\ref{105},\ref{W}) using the map between $\phi$ and $\Phi$ fields. 
	To obtain the gauge couplings we adopt the iterative Noether's scheme. According to this scheme the first correction to the lagrangian is given by,
	\begin{equation}
		\mathcal{L}_1=-j^iA_i=-i\Big(\Phi^2(\Phi^*\partial_i\partial_j\Phi^*-\partial_i\Phi^*\partial_j\Phi^*)-{\Phi^*}^2(\Phi\partial_i\partial_j\Phi-\partial_i\Phi\partial_j\Phi)\Big)\partial_jA_i\label{h3}
	\end{equation}
	Since derivatives (on the matter field) persist, there will be another correction. This is given by, 
	\begin{equation}
		j^i=\partial_j\Big(|\Phi|^4(\partial_jA_i+\partial_iA_j)\Big)\label{h4}
	\end{equation}
	where we have used $\mathcal{L}_1$ (\ref{h3}) in (\ref{h1}) to get this current. Then the contribution to the lagrangian is given by,
	\begin{equation}
		\mathcal{L}_2=-j^iA_i=\frac{1}{4}|\Phi|^4(\partial_iA_j+\partial_jA_i)^2\label{h5}
	\end{equation}
	so that the current is reproduced by,
	\begin{equation}
		j^i=-\frac{\delta}{\delta A_i}\int \mathcal{L}_2\label{h6}
	\end{equation}
	Since there are no further derivatives on $\Phi$, the iteration terminates and the final form of the gauged theory is given by,
	\begin{align}
		\mathcal{L}&=\mathcal{L}_0+\mathcal{L}_1+\mathcal{L}_2\nonumber\\
		&=|\Phi\partial_i\partial_j\Phi-\partial_i\Phi\partial_j\Phi|^2+\frac{i}{2}\Big(\Phi^2(\Phi^*\partial_i\partial_j\Phi^*-\partial_i\Phi^*\partial_j\Phi^*)-{\Phi^*}^2(\Phi\partial_i\partial_j\Phi-\partial_i\Phi\partial_j\Phi)\Big)(\partial_iA_j+\partial_jA_i)\nonumber\\
		&+\frac{1}{4}|\Phi|^4(\partial_iA_j+\partial_jA_i)^2\label{h7}
	\end{align}
	which exactly matches with the form (\ref{h51}).\\
	It is also possible to recast the above analysis in terms of the tensor field defined in (\ref{a22}). In that case the first correction (\ref{h3}) is expressed as,
	\begin{equation}
		\mathcal{L}_1=-i\Big(\Phi^2(\Phi^*\partial_i\partial_j\Phi^*-\partial_i\Phi^*\partial_j\Phi^*)-{\Phi^*}^2(\Phi\partial_i\partial_j\Phi-\partial_i\Phi\partial_j\Phi)\Big)A_{ij}=-j^{ij}A_{ij}\label{h8}
	\end{equation}
	where the current $j^{ij}$ coupling to $A_{ij}$ is defined in (\ref{h2}). Pursuing this line of analysis the new (tensor) current obtained from the first iteration is given by,
	\begin{equation}
		j^{ij}=-2|\Phi|^4A_{ij}\label{h9}
	\end{equation}
	which is consistent with (\ref{h4}). The new contribution to the lagrangian now follows,
	\begin{equation}
		\mathcal{L}_2=|\Phi|^4A_{ij}^2=-j^{ij}A_{ij}\label{h10}
	\end{equation}
	so that the current (\ref{h9}) is reproduced by,
	\begin{equation}
		j^{ij}=- \frac{\delta}{\delta A_{ij}}\int\mathcal{L}_2\label{h11}
	\end{equation}
	Note that the structure (\ref{h10}) is consistent with (\ref{h5}) on using the relation (\ref{a22}). Thus the final lagrangian takes the form,
	\begin{align}
		\mathcal{L}&=\mathcal{L}_0+\mathcal{L}_1+\mathcal{L}_2\nonumber\\
		&=|\Phi\partial_i\partial_j\Phi-\partial_i\Phi\partial_j\Phi|^2-i\Big[\Phi^2(\Phi^*\partial_i\partial_j\Phi^*-\partial_i\Phi^*\partial_j\Phi^*)-{\Phi^*}^2(\Phi\partial_i\partial_j\Phi-\partial_i\Phi\partial_j\Phi)\Big]A_{ij}+|\Phi|^4A_{ij}^2\label{h12}
	\end{align}
	which is the dual version of (\ref{h7}) and is same as (\ref{h52}).
	\subsection{Full interacting theory in dual description}
	As we have shown in section-4 that, for complex scalar theories with linear space dependent shift symmetry, the gauging can be done, either by making use of tensor gauge fields or by using usual vector gauge fields. We have shown that this is a consequence of the functional relationship between the vector and higher rank tensor currents. However, this was only discussed algebraically as no gauge field dynamics was involved in the earlier sections. In this section we shall discover that the mapping is even valid at the dynamical level. By taking the full interacting lagrangian where the gauge fields, either vector or tensor, are also dynamical, descriptions of these theories are possible not only at the lagranian level but also at the hamiltonian level.
	\subsubsection{Lagrangian and hamiltonian analysis in (2+1) dimensional space-time}
	The interacting theory can be presented in two distinct ways. As a symmetric tensor theory or a mapped vector theory using (\ref{a22}) in the tensor theory. We use the mapping in two different ways: (1) we write down a lagrangian in terms of vector fields from the tensor theory lagrangian using the map and then find the equation of motions of the vector fields. (2) On the other hand we can just use the map at the level of equation of motion of the tensor theory itself to find a set of vector field equations of motion and see that in both cases the vector equations of motion match. 
	\\We first consider the (2+1) dimensional case. In (2+1) dimension the Maxwell theory for U(1) vector fields modifies from that of (3+1) dimensional theory. The electric field is a 2 dimensional vector $E_i=-F_{oi}=\partial_iA_0-\partial_0A_i$ and the magnetic field is a scalar  $B=\epsilon_{ij}\partial_iA_j$.\\\\
	\textbf{Theory with second rank tensor gauge fields}\\
	Now let us write a (2+1) dimensional theory in terms of tensorial gauge fields. For a tensorial theory in 2+1 dimension, the electric field is taken as a rank-2 symmetric tensor, while the magnetic field becomes a vector. Let us first define the electric and magnetic fields for this tensor theory in terms of a symmetric rank-2 tensor gauge field $A_{ij}$ as, 
	\begin{equation}
		E_{ij}=\partial_0A_{ij}+\partial_i\partial_jA_0,\quad B_k=\epsilon_{ij}\partial_iA_{jk}\label{b9}
	\end{equation}
	One can check that both these fields are gauge invariant. The lagrangian for the pure gauge field is written, 
	\begin{align}
		\mathcal{L}_0&=\frac{1}{2}(E_{ij}^2-B_i^2)\label{b7}\\
		&=\frac{1}{2}\Big[(\partial_i\partial_jA_0)^2+(\partial_0A_{ij})^2+2(\partial_0A_{ij})(\partial_i\partial_jA_0)\Big]-\frac{1}{2}(\partial_mA_{nk}-\partial_nA_{mk})\partial_mA_{nk}\label{a60}
	\end{align}
	The complete theory is now given by,
	\begin{align}
		\mathcal{L}&=\mathcal{L}_0+\frac{i}{2}\Phi^*\overleftrightarrow{\partial}_0\Phi-\Phi^*\Phi A_0+|\Phi\partial_i\partial_j\Phi-\partial_i\Phi\partial_j\Phi|^2\nonumber\\
		&-i\Big[\Phi^2(\Phi^*\partial_i\partial_j\Phi^*-\partial_i\Phi^*\partial_j\Phi^*)-{\Phi^*}^2(\Phi\partial_i\partial_j\Phi-\partial_i\Phi\partial_j\Phi)\Big]A_{ij}+|\Phi|^4A_{ij}^2\nonumber\\
		&=\mathcal{L}_0+\frac{i}{2}\Phi^*\overleftrightarrow{\partial}_0\Phi-\Phi^*\Phi A_0+|\Phi\partial_i\partial_j\Phi-\partial_i\Phi\partial_j\Phi|^2-j_{ij}A_{ij}
	\end{align}
	where, 
	\begin{equation}
		j_{ij}=i\Big[\Phi^2(\Phi^*\partial_i\partial_j\Phi^*-\partial_i\Phi\partial_j\Phi)-{\Phi^*}^2(\Phi\partial_i\partial_j\Phi-\partial_i\Phi\partial_j\Phi)\Big]-|\Phi|^4A_{ij}\label{t2}
	\end{equation}
	The equation of motion for $A_0$ is given by
	\begin{equation}
		\partial_i\partial_jE_{ij}=-\Phi^*\Phi\label{r6}
	\end{equation}
	while that of the gauge field $A_{ij}$ is given by
	\begin{equation}
		\partial_0E_{ij}+\frac{1}{2}\partial_k(\epsilon_{ik}B_j+\epsilon_{jk}B_i)=-j_{ij}\label{r10}
	\end{equation}
	Lastly, the equation of motion of the scalar field $\Phi$ is given by 
	\begin{align}
		&-(i\partial_0+A_0)\Phi^*+4(\partial_i\partial_j\Phi)\mathcal{C}_{ij}+4(\partial_i\Phi)\partial_j\mathcal{C}_{ij}+\Phi\partial_i\partial_j\mathcal{C}_{ij}-i\Phi\mathcal{C}_{ij}A_{ij}+4i{\Phi^*}^2(\partial_i\partial_j\Phi)A_{ij}
		\nonumber\\
		&+2{\Phi^*}^2\Phi A_{ij}^2-\partial_i\Big((\Phi^*)^2A_{ij}\Big)(\partial_j\Phi)-\Phi\partial_i\partial_j\Big((\Phi^*)^2A_{ij}\Big)=0 \label{13}
	\end{align}
	where 
	\begin{equation}
		\mathcal{C}_{ij}=\Phi^*\partial_i\partial_j\Phi^*-\partial_i\Phi^*\partial_j\Phi^*\label{c}
	\end{equation}
	Now let us compute the Hamilton's equations of motion for the electric and magnetic fields and deduce the respective Maxwell equations for the tensor theory. 
	From (\ref{a60}),  one can compute the canonical momenta corresponding to $A_0$ and $A_{ij}$ as
	\begin{equation}
		\pi_0=\frac{\partial\mathcal{L}}{\partial\dot{A}^{0}}=0,\qquad \pi_{ij}=\frac{\partial\mathcal{L}}{\partial\dot{A}^{ij}}=\dot{A}_{ij}+\partial_i\partial_jA_0=E_{ij}\label{a33}
	\end{equation}
	and the corresponding canonical Hamiltonian as
	\begin{equation}
		H_c=\frac{1}{2}(E_{ij}^2+B_k^2)+...\label{a53}
	\end{equation}
	where the ellipses denote terms containing $\Phi$ and the interaction terms.
	The basic Poission relation between the gauge field $A_{ij}$ and its canonical momenta $E_{ij}$ is given by
	\begin{equation}
		\{A_{ij}(\vec{x}),A_{kl}(\vec{y})\}=0=\{E_{ij}(\vec{x}),E_{kl}(\vec{y})\};\quad	\{A_{ij}(\vec{x}),E_{kl}(\vec{y})\}=\frac{1}{2}(\delta_{ik}\delta_{jl}+\delta_{il}\delta_{jk})\delta^2(\vec{x}-\vec{y})\label{a49}
	\end{equation}
	while that between the physical variables (electric and magnetic fields) are given by
	\begin{equation}
		\{B(\vec{x}),B(\vec{y})\}=0,\qquad	\{E_{ij}(x),B_k(y)\}=\frac{1}{2}\partial_m^y(\epsilon_{mi}\delta_{jk}+\epsilon_{mj}\delta_{ik})\delta^2(\vec{x}-\vec{y})    \label{a51}
	\end{equation}
	Now using (\ref{a49}) we compute the time derivative of the electric field as
	\begin{equation}
		\partial_0E_{ij}=\{E_{ij},H_c\}=\frac{1}{2}(\epsilon_{ki}\partial_kB_j+\epsilon_{kj}\partial_kB_i)-j_{ij}\label{a54}
	\end{equation}
	This corresponds to Faraday's law for tensor theory in (2+1) dimension. Similarly the equation of motion of $B_i$ is given by
	\begin{equation}
		\partial_0B_i=\{B_i,H_c\}=-\epsilon_{jk}\partial_jE_{ki}\label{a56}
	\end{equation}
	which corresponds to Ampere's law for tensor theory in (2+1) dimensions.\\\\
\textbf{Theory with vector gauge field}\\\\
	We now use the map (\ref{a22}) between the rank-2 symmetric tensor gauge field and usual vector field to write down a lagrangian in terms of vector fields. Then we shall find out the equations of motion for the vector fields.\\
	Using (\ref{a22}) in the lagrangian (\ref{a60}), the vector gauge field lagrangian for (2+1) dimensional theory is given by
	\begin{equation}
		\mathcal{L}_0=\frac{1}{4}\Big[2(\partial_i\partial_jA_0)^2+(\partial_0\partial_iA_j)^2-4(\partial_i\partial_jA_0)(\partial_i\partial_0A_j)+(\partial_i\partial_0A_j)(\partial_j\partial_0A_i)\Big]-\frac{1}{8}(\epsilon_{jk}\partial_i\partial_jA_k)^2\label{b10}
	\end{equation}
	The full lagrangian containing scalar field part and the interaction part will be given by
	\begin{align}
		&\mathcal{L}=\mathcal{L}_0+\frac{i}{2}\Phi^*\overleftrightarrow{\partial}_0\Phi-\Phi^*\Phi A_0+|\Phi\partial_i\partial_j\Phi-\partial_i\Phi\partial_j\Phi|^2\nonumber-j_iA_i\label{b5}
	\end{align} 
	where (\ref{B2}) has been used to recast the interaction term in favour of vector quantities.
	Now we can compute the equation of motion of the fields from this lagrangian. First, the equation of motion of $A_0$ field is given by
	\begin{equation}
		\partial^2\partial_iE_{i}=-\Phi^*\Phi\label{b6}
	\end{equation}
	which corresponds to the Gauss constraint, where $E_i$ is defined as before $E_i=-F_{oi}$, However note that it is not the canonical momenta corresponding to $A_i$. This will be shown in details in the Hamiltonian analysis. The equation of motion for $A_i$ is given by
	\begin{equation}
		\frac{1}{2}\partial^2(\partial_0E_{i}-\frac{1}{2}\epsilon_{ij}\partial_jB)+\frac{1}{2}\partial_0\partial_i\partial_jE_j=-j_i \label{a59}   
	\end{equation}
	where 
	\begin{equation}
		j_i=i\partial_j[\Phi^2(\Phi^*\partial_i\partial_j\Phi^*-\partial_i\Phi\partial_j\Phi)-{\Phi^*}^2(\Phi\partial_i\partial_j\Phi-\partial_i\Phi\partial_j\Phi)] +\partial_j[|\Phi|^4(\partial_iA_j+\partial_jA_i)]\label{r3}
	\end{equation}
	is the vector current for the theory \cite{dual} and is related to a tensor current $j_{ij}$ as
	\begin{equation}
		j_i=\partial_jj_{ij};\qquad j_{ij}=[i\Phi^2(\Phi^*\partial_i\partial_j\Phi^*-\partial_i\Phi\partial_j\Phi)-i{\Phi^*}^2(\Phi\partial_i\partial_j\Phi-\partial_i\Phi\partial_j\Phi)+|\Phi|^4(\partial_iA_j+\partial_jA_i)]\label{j}
	\end{equation}
	The equation of motion of $\Phi$ is essentially the same as (\ref{13}), because the interaction lagrangian is not changed, apart from the fact that $A_{ij}$ will now be replaced by $-\frac{1}{2}(\partial_iA_j+\partial_jA_i)$ in this picture. \\\\We shall now discuss about the hamiltonian analysis of the vector theory. From  (\ref{b10}), one can find out the canonical momenta for the vector fields as
	\begin{equation}
		\pi_0=\frac{\partial\mathcal{L}}{\partial\dot{A^0}}=0\,;\qquad	\pi_i=\frac{\partial\mathcal{L}}{\partial\dot{A^i}}=\frac{1}{2}(\delta_{ij}\partial^2+\partial_i\partial_j) E_j\label{162}
	\end{equation}
	The canonical Hamiltonian for the theory can be written using Legendre transformation as 
	\begin{equation}
		H_c=\frac{1}{2\partial^4}(\partial_i\pi_j)^2 +\frac{1}{8}(\partial_i B)^2+...\label{a46}
	\end{equation}
	where the ellipses mean terms containing scalar fields and interactions.\\The Poission algebra between the canonically conjugate fields $A_i$ and $\pi_i$ are given by
	\begin{equation}
		\{A_i(\vec{x}),A_j(\vec{y})\}=\{\pi_i(\vec{x}),\pi_j(\vec{y})\}=0;\qquad \{A_{i}(\vec{x}),\pi_{j}(\vec{y})\}=\delta_{ij}\delta^2 (\vec{x}-\vec{y})\label{a10}
	\end{equation} 
	while that between the electric and magnetic fields are,
	\begin{equation}
		\{\pi_i,\pi_j\}=0=\{B,B\},\quad \{B(\vec{x}),\pi_j(\vec{y})\}=\epsilon_{kj}\partial_k^x\delta^2 (\vec{x}-\vec{y})
	\end{equation}
	Using this algebra, we can derive the equation of motion of $E_i$  and $B$ from the above Hamiltonian as
	\begin{align}
		\partial_0\pi_i=\{\pi_i,H_c\}=\frac{\partial^2}{4}\epsilon_{ji}\partial_j &B-j_i\nonumber\\
		\Rightarrow	\frac{\partial^2}{2}(\partial_0E_i-\frac{1}{2}\epsilon_{ij}\partial_jB)+\frac{1}{2}\partial_0\partial_i\partial_jE_j&=-j_i\label{a47}
	\end{align}
	and
	\begin{equation}
		\partial_0 B=\{B,H_c\}=\epsilon_{ji}\partial_jE_i\label{a48}
	\end{equation}
	\vspace{5pt}
	\textbf{Comparative study}\\
	We cannot compare the equations of motion derived in two scenarios at this stage, as the first case provides equation of motion for tensor fields whereas the second case considers vector fields.  The first step is to find a map connecting the physical gauge invariant variables (electric and magnetic fields) in the two pictures. Exploiting the basic map (\ref{a22}) in (\ref{b9}), the following identifications are obtained,
	\begin{equation}
		E_{ij}=\frac{1}{2}(\partial_iE_j+\partial_jE_i),\qquad B_k=\epsilon_{ij}\partial_iA_{jk}=-\frac{1}{2}\partial_k B\label{a50}
	\end{equation}
	where $E_i$ and $B$ were defined earlier.\\
	The Gauss constraint may now be compared. From (\ref{r6}) we can write using (\ref{a50}) ,
	\begin{equation}
		-\Phi^*\Phi=\partial_i\partial_jE_{ij}=\frac{1}{2}\partial_i\partial_j(\partial_iE_j+\partial_jE_i)=\partial^2\partial_iE_i\label{p61}
	\end{equation}
	which is same as (\ref{b6}) but now derived directly from the tensor field equations using the mapping.\\ Using the map (\ref{a50}) in (\ref{r10}), we obtain,
	\begin{equation*}
\frac{1}{2}\partial_j[\partial_0E_{i}-\frac{1}{2}\epsilon_{im}\partial_j\partial_mB]+\frac{1}{2}\partial_0\partial_iE_j= -j_{ij}
	\end{equation*}
	Now using $\partial_j$ on the above equation from left we get
		\begin{equation}
		 	\Rightarrow\frac{1}{2}(\partial^2\partial_0E_i+\partial_0\partial_i\partial_jE_j)-\frac{1}{4}\partial^2 \epsilon_{ij}\partial_j B=-j_i\label{p60}
	\end{equation}
	and (\ref{a47}) is reproduced. Moreover, it is feasible to do a consistency check by acting a $\partial_i$ on (\ref{p60}) and see using (\ref{b6}) that it reproduces the usual current conservation law for vector fields,
	\begin{equation}
	    \partial_0(\Phi^*\Phi)-\partial_ij_i=\partial_0j_0-\partial_ij_i=0
	\end{equation}
	Let us now see the comparison at the hamiltonian level.
	 It is seen that one can also arrive at (\ref{a46}) from the tensor theory Hamiltonian (\ref{a53}), using the map (\ref{a50}).
	Now the tensorial Ampere's law (\ref{a56}) can be mapped to the following equation using (\ref{a50}).  
	\begin{equation}
		\partial_0\partial_i B=\epsilon_{jk}\partial_i\partial_jE_k\label{a57}
	\end{equation}
	The solution of the above equation gives us the corresponding Ampere's law for vector theory (\ref{a48}).\\
	The above comparison can be captured using the following chart.\\
	\begin{tabular}{ | m{1.5cm} | m{4.0cm}| m{5.5cm}|m{3.8cm}|} 
	\hline
	&Basic gauge field& Electric field &Magnetic field\\
	\hline
	Fields& $A_i,\,\,A_{ij}$& $E_i = \partial_iA_0-\partial_0A_i,\,\,E_{ij}$& $B = \epsilon_{jk}\partial_jA_k,\,\,B_i$\\
	\hline
	Derived Map& $A_{ij} =-\frac{1}{2}(\partial_iA_j + \partial_jA_i) $& $E_{ij} =
\frac{1}{2}	(\partial_iE_j + \partial_jE_i)$& $B_i=-\frac{1}{2}	\partial_iB$\\
\hline
\end{tabular}\\
\begin{center}
	Table I : Comparison of various fields in (2+1) dimension for linear shift symmetric theory
\end{center}	
This table shows how the basic fields in the rank two theory $(A_{ij},E_{ij},B_i)$ are related to the corresponding variables $(A_i,E_i,B)$ in the vector theory.To complete the picture we provide another table.\\ \\
	\begin{tabular}{ | m{2.0cm} | m{3.5cm}| m{6.3cm}|m{3.8cm}|} 
		\hline
		& Gauss law & Faraday's law & Ampere's law\\
		\hline
		Tensor theory & $\partial_i\partial_jE_{ij}+\Phi^*\Phi=0$&$		\partial_0E_{ij}+\frac{1}{2}(\epsilon_{ik}\partial_kB_j+\epsilon_{jk}\partial_kB_i)=-j_{ij}$&$\partial_0B_{i}=-\epsilon_{jk}\partial_jE_{ki}$\\ 
				\hline
		Using (\ref{a50}) in tensor theory &
		$\partial^2\partial_iE_i+\Phi^*\Phi=0$&$   \frac{1}{2}\partial_j[\partial_0E_{i}-\frac{1}{2}\epsilon_{im}\partial_j\partial_mB]+\frac{1}{2}\partial_0\partial_iE_j= -j_{ij}$& $\partial_i\partial_0B= \epsilon_{jk}\partial_i\partial_jE_{k}$\\
		\hline
		& & Using $\partial_j$ on above equation &Acting with $\frac{1}{\partial_i}$ on above eqn\\
		\hline

		Vector theory&$\partial^2\partial_iE_i+\Phi^*\Phi=0$& $\frac{1}{2}\partial^2[\partial_0E_{i}-\frac{1}{2}\epsilon_{ij}\partial_jB]+\frac{1}{2}\partial_0\partial_i\partial_jE_j= -j_i $& $\partial_0B=\epsilon_{jk}\partial_jE_k$\\ 
		\hline
	\end{tabular}

	\begin{center}Table II : Comparison of various equations of motion in (2+1) dimension for linear shift
		symmetric theory
	\end{center}
To use this table one has to adopt the following approach. The first and last rows give the results obtained from the tensor coupling and vector coupling, respectively. The results for the vector case are reproduced from the tensor theory using the maps given in table I. It is either obtained directly (Gauss law) or by applying suitable derivative operators (Faraday's and Ampere's law). It is expected as  Faraday's and Ampere's law for tensor and vector theories differ by one rank whereas  Gauss law being a scalar equation in both  cases is reproduced automatically.\\\\
The different descriptions of gauge theories presented here need some elaboration. Starting from a rank-2 tensor theory \eqref{a60}, the vector theory \eqref{b10} was constructed using the map \eqref{a22}. These theories were connected in the sense that the physical equations- Gauss law, Faraday's law, and Ampere's law - were identical, irrespective of the fact that these were directly obtained from the vector theory or derived from the tensor theory exploiting suitable maps.\\\\
It should be pointed out that the map \eqref{a22} allows a passage from the tensor theory to the vector theory but not the other way around. All correspondences have to be interpreted in this way, either in the action formulation or in the equation of motion.  While the tensor Faraday's equations pass to the vector ones, the converse does not hold. In passing from the vector to the tensor equations, the transverse components remain undetermined. This is bacause, there is a mismatch in the number of degrees of freedom of the gauge fields. To see this more elaborately, we recall that the two degrees of freedom of the vector field are the transverse and longitudinal components, while the three degrees of freedom of the tensor field can be decomposed as a fully transverse component, a partially transverse component, and a longitudinal component. While the longitudinal components of the two gauge fields may be mapped, this is not true for the transverse components. \\\\
The map \eqref{a22} was used since the primary motivation was to discuss a connection among various gauge theories, apart from the fact that it occurs naturally in the present analysis and has also been mentioned earlier in the literature \cite{seiberg,dual}. Similar discussion is also valid for the subsequent sections when we discuss about 3+1 dimensional case and quadratic shift symmetric theory. 
	\subsubsection{Lagrangian and hamiltonian analysis in (3+1) dimensional space-time}
	\textbf{Theory with second rank tensor gauge field}\\\\
	Here we shall write down the gauge field lagrangian for tensorial gauge fields in (3+1) dimension, which also has to be gauge invariant. So, first let us write the gauge invariant electric and magnetic field tensor \cite{pretko,dual} corresponding to $A_{ij}$ as, 
	\begin{align}
		&E_{ij}=F_{0ij}=\partial_i\partial_jA_0+\partial_0A_{ij}\label{a8}\\
		&B_{ij}=\partial_kF_{kij}=\partial_k(\partial_jA_{ik}-\partial_i A_{jk})\label{a9}
	\end{align}
	Note that the magnetic field is now changed from the (2+1) dimensional case, which is now a two rank anti-symmetric tensor.\\
	The pure gauge field lagrangian, in analogy with the Maxwell example, is given by
	\begin{align}
		\mathcal{L}_{0}&=\frac{1}{2}(E_{ij}^2-B_{ij}^2)\nonumber\\
		&=\frac{1}{2}\Big[(\partial_i\partial_jA_0)^2+(\partial_0A_{ij})^2\Big]+(\partial_0A_{ij})(\partial_i\partial_jA_0)-(\partial_i\partial_kA_{kj})^2+(\partial_i\partial_kA_{kj})(\partial_j\partial_lA_{li})\label{a20}
	\end{align}
	We can construct the full lagrangian in terms of $A_{ij}$ fields as
	\begin{equation}
		\mathcal{L}= \mathcal{L}_{0}+\frac{i}{2}\Phi^*\overleftrightarrow{\partial}_0\Phi- \Phi^*\Phi A_0+|\Phi\partial_i\partial_j\Phi-\partial_i\Phi\partial_j\Phi|^2-j_{ij}A_{ij}\label{a34}
	\end{equation}
	Note that terms involving $A_0$ and pure matter and matter-gauge interaction will remain unchanged from the previous case of (2+1) dimension. The tensor current $j_{ij}$ is already defined before in (\ref{t2}).\\
	The equation of motion for $A_0$ is given by
	\begin{equation}
		\partial_i\partial_jE_{ij}+\Phi^*\Phi=0\label{l}
	\end{equation}
	which corresponds to the Gauss law, and the equation of motion of $A_{ij}$ as given by
	\begin{equation}
		\partial_0E_{ij}-\frac{1}{2}\partial_k(\partial_iB_{kj}+\partial_j B_{ki})=-j_{ij} \label{16} 
	\end{equation}
	The equation of motion of $\Phi$ is same as that of (\ref{13}).\\\\
	One can compute the canonical momenta from lagrangian (\ref{a20}) corresponding to $A_0$ and $A_{ij}$ as,
	\begin{equation}
		\pi_0=\frac{\partial\mathcal{L}}{\partial\dot{A}^{0}}=0,\qquad \pi_{ij}=\frac{\partial\mathcal{L}}{\partial\dot{A}^{ij}}=\dot{A}_{ij}+\partial_i\partial_jA_0=E_{ij}\label{aa33}
	\end{equation}
	So the canonical Hamiltonian can be written as
	\begin{equation}
		H_C= \frac{1}{2}(E_{ij}^2+B_{ij}^2)+...\label{a37}
	\end{equation}
	The ellipses denote terms containing the pure matter sector and interactions.\\
	Now let us compute the equation of motion of the tensorial electric field $E_{ij}$ and magnetic field $B_{ij}$ directly using the canonical Hamiltonian $H_c$ (\ref{a37}). So the time derivative of the electric field tensor is given by
	\begin{align}
		\partial_0E_{ij}(x)&=\{E_{ij},H_C\}=\int d^3y [\{E_{ij}(x),B_{mn}(y)\}B_{mn}(y)+\{E_{ij}(x),A_{kl}(y)j_{kl}(y)\}]\nonumber\\
		&= \frac{1}{2}(\partial_k\partial_iB_{kj}+\partial_k\partial_jB_{ki})(x)-j_{ij}(x)\label{a58}
	\end{align}
	where we have made use of (\ref{a49}). One can identify the right hand side of the above relation as curl of the tensor field $B_{ij}$. So the above equation (\ref{a58}) represents the counterpart of Faraday's law for tensorial gauge theory. Similarly we can derive the time derivative of the magnetic field, given by
	\begin{align}
		\partial_0B_{ij}(x)&=\{B_{ij},H_C\}=\int d^3y \{B_{ij}(x),E_{mn}(y)\}E_{mn}(y)\nonumber\\
		&= -(\partial_k\partial_iE_{kj}-\partial_k\partial_jE_{ki})\label{a43}
	\end{align}
	which corresponds to the Ampere's law studied in ususal Maxwell theory.\\\\
	\textbf{Theory with vector gauge fields}\\\\
	We now use the map (\ref{a22}) between the rank-2 symmetric tensor gauge field and usual vector field to write down a a lagrangian in terms of vector fields just like we did in (2+1) dimensional theory. Then we shall find out the equation of motions for the vector fields.\\
	Using (\ref{a22}) in the lagrangian (\ref{a20}) the gauge field lagrangian for (3+1) dimensional theory is given by
	\begin{equation}
		\mathcal{L}_0=\frac{1}{4}\Big[2(\partial_i\partial_jA_0)^2+(\partial_0\partial_iA_j)^2-4(\partial_i\partial_jA_0)(\partial_i\partial_0A_j)+(\partial_i\partial_0A_j)(\partial_j\partial_0A_i)\Big]-\frac{1}{8}(\partial^2(\partial_iA_j-\partial_jA_i))^2\label{bb10}
	\end{equation}
	The full lagrangian containing scalar field part and the interaction part will be given by
	\begin{align}
		&\mathcal{L}=\mathcal{L}_0+\frac{i}{2}\Phi^*\overleftrightarrow{\partial}_0\Phi-\Phi^*\Phi A_0+|\Phi\partial_i\partial_j\Phi-\partial_i\Phi\partial_j\Phi|^2\nonumber-j_iA_i\label{b5}
	\end{align} 
	Now we can compute the equation of motion of the fields from this lagrangian. First, the equation of motion of $A_0$ field is given by
	\begin{equation}
		\partial^2\partial_iE_{i}=-\Phi^*\Phi\label{bb6}
	\end{equation}
	which corresponds to the Gauss constraint. Now the equation of motion corresponding to $A_i$ is given by
	\begin{equation}
		\frac{\partial^2}{2}\Big[\partial_0E_i-\frac{\partial^2}{2}\epsilon_{ijk}\partial_jB_k\Big]+\frac{1}{2}\partial_0\partial_i\partial_jE_j=-j_i\label{p90}   
	\end{equation}\\
	One can check that the canonical momenta corresponding to the field $A_{\mu}$ obtained from the lagrangian (\ref{bb10}) are given by the same expression as (\ref{162}) as there is no change in the electric field.\\ 
	The canonical hamiltonian corresponding to the theory is written 
	as
	\begin{equation}
		H_c=\frac{1}{2\partial^4}(\partial_i\pi_j)^2+\frac{1}{8}(\epsilon_{ijk}\partial^2 B_k)^2+....\label{hc2}
	\end{equation}
The ellipses denote the terms containing scalar fields and interaction terms.\\
	Using the basic Poission relation between gauge field $A$ and its canonical momentum $\pi$ given in (\ref{a10}), we can write down the equation of motion of $E_i$ from the above Hamiltonian as
	\begin{align}
		&\partial_0\pi_{i}(x)=\{\pi_{i},H_C\}\nonumber\\
		\Rightarrow & \frac{\partial^2}{2}[\partial_0 E_i-\frac{\partial^2}{2}\epsilon_{ikj}\partial_kB_j]+\frac{1}{2}\partial_0\partial_i\partial_jE_j=-j_i\label{b12}
	\end{align}
	Similarly the equation of motion for the field $B_i$ is given by
	\begin{equation}
		\partial_0B_{i}(x)=\{B_{i},H_C\}= -\epsilon_{ijk}(\partial_jE_k-\partial_kE_j)\label{b13}
	\end{equation}
 \newpage
\textbf{Comparative study}\\\\
	Now to compare the equations of motion in both pictures we must bring them at the same level.  We cannot compare two pictures as the former equations of motion are of tensor fields whereas the later are equations of motion of vector fields. 
	So first we find a map between the physical variables (electric and magnetic field) of both pictures. The physical variables of tensor theory can be expressed in favour of the Maxwell field $E_i$ and $B_i$ as
	\begin{align}
		E_{ij}=&\frac{1}{2}(\partial_iF_{j0}+\partial_jF_{i0})=\frac{1}{2}(\partial_iE_j+\partial_jE_i);\qquad E_i=-F_{0i}\label{b1}\\
		B_{ij}=&\frac{1}{2}\partial^2F_{ij}=\frac{1}{2}\,\epsilon_{ijk}\,\partial^2B_k\label{b2}
	\end{align}
Using (\ref{b1}), we can write from the tensor equation corresponding to $A_0$ (\ref{l}) as
	\begin{equation}
		-\Phi^*\Phi=\partial_i\partial_jE_{ij}=\partial^2\partial_iE_i
	\end{equation}
	which is same as we derived in (\ref{bb6}). We can also map the equation of motion of tensor field $A_{ij}$ (\ref{16}) to a vector field equation of motion. From (\ref{16}) we can write 
$$ \frac{\partial_j}{2}[\partial_0 E_i-\frac{\partial^2}{2}\epsilon_{ikl}\partial_kB_l]+\frac{1}{2}\partial_0\partial_iE_j=-j_{ij}$$
	By acting $\partial_j$ operator from left on the above and using $j_i=\partial_jj_{ij}$, we reproduce the same equation (\ref{p90}).  \\Now let us take the Ampere's law for tensor theory (\ref{a43}) and use the map (\ref{b1},\ref{b2}) to get
	\begin{equation}
		\partial^2\Big[\frac{1}{2}\epsilon_{ijk}\partial_0B_k+(\partial_iE_j-\partial_jE_i)\Big]=0
	\end{equation}
	By acting $\epsilon_{ijm}$ on the above equation we get
	\begin{equation}
		\partial^2\Big[\partial_0B_m+\epsilon_{ijm}(\partial_iE_j-\partial_jE_i)\Big]=0 
	\end{equation}
	the solution of which is the vector equation of motion of $B_m$ given by (\ref{b13}) from the vector theory. 
	Now the comparison can be illustrated from the following chart:\\
	
	\begin{tabular}{ | m{1.5cm} | m{3.9cm}| m{6.0cm} |m{4.0cm} |}
	\hline
		&Gauge field& Electric field &Magnetic field\\
	\hline
	Fields& $A_i,\,\,A_{ij}$& $E_i = \partial_iA_0-\partial_0A_i,\,\,E_{ij}$& $B_i = \epsilon_{ijk}\partial_jA_k,\,\,B_{ij}$\\
	\hline
	Mapping& $A_{ij} =-\frac{1}{2}(\partial_iA_j + \partial_jA_i) $& $E_{ij} =
	\frac{1}{2}	(\partial_iE_j + \partial_jE_i)$& $B_{ij}=\frac{1}{2}\epsilon_{ijk}\partial^2B_k$\\
	\hline
\end{tabular}	\\
\begin{center}
	Table III : Comparison of various fields in (3+1) dimension for linear shift symmetric theory
\end{center}
\begin{tabular}{ | m{1.5cm} | m{3.4cm}| m{6.5cm} |m{4.9cm} |}
		\hline
		& Gauss law& Faraday's law&Ampere's law \\
		\hline
		Tensor theory & $\partial_i\partial_jE_{ij}+\Phi^*\Phi=0$&$		\partial_0E_{ij}-\frac{1}{2}\partial_k(\partial_iB_{kj}+\partial_j B_{ki})=-j_{ij}$&$\partial_0B_{ij}+\partial_k(\partial_iE_{kj}-\partial_jE_{ki})=0$\\ 
		\hline
		Using map (\ref{b1},\ref{b2}) in tensor theory & $\partial^2\partial_iE_{i}+\Phi^*\Phi=0$ & $   \frac{\partial_j}{2}[\partial_0 E_i-\frac{\partial^2}{2}\epsilon_{ikl}\partial_kB_l]+\frac{1}{2}\partial_0\partial_iE_j=-j_{ij}$&$\partial^2[\frac{1}{2}\epsilon_{ijk}\partial_0B_k+(\partial_iE_j-\partial_jE_i)]=0$\\ 
		\hline
		&& Acting $\partial_j$ on the above equation
	& Acting $\frac{\epsilon_{ijm}}{\partial^2}$ on the above equation  \\
		\hline
			Vector theory&$\partial^2\partial_iE_{i}+\Phi^*\Phi=0$& $\frac{\partial^2}{2}[\partial_0 E_i-\frac{\partial^2}{2}\epsilon_{ikj}\partial_kB_j]+\frac{1}{2}\partial_0\partial_i\partial_jE_j=-j_i$ &$\partial_0B_m+\epsilon_{ijm}(\partial_iE_j-\partial_jE_i)=0$ \\ 
			\hline
	\end{tabular}
\begin{center}
Table IV : Comparison of various equations of motion in (3+1) dimension for linear shift symmetric theory
\end{center}
Above table should be interpreted in the same way as was described after table II. 
	\section{Dual description of theory with space-dependent quadratic shift symmetry}
	The interplay between a tensor formulation and the usual vector formulation is generic, not confined to the linear shift symmetry. Infact it is more general. In this sub-section we show that the non-standard gauging prescription for a lagrangian having quadratic space dependent shift symmetry involves a rank-3 tensor gauge field which is either described by the usual minimal prescription based on a vector gauge field or by a rank-2 tensor field.\\ \\
	Quadratic space dependent shift transformation on the complex scalar field can be written as,
	\begin{equation}
		\Phi(x)\to e^{iw(x)}\Phi(x)\quad ;\quad w(x)=c+c_ix^i+\frac{1}{2}c_{ij}x^ix^j,\quad c,c_i,c_{ij}=\textrm{constant parameters}\label{1}	\end{equation}
A specific form for	the matter lagrangian having the above symmetry (\ref{1}) is given by \cite{hoyos}
	\begin{align}
		\mathcal{L}&=u\Big|{\Phi^*}^3\Big(\Phi^2\partial_i\partial_j\partial_k\Phi+2\partial_i\Phi\partial_j\Phi\partial_k\Phi-\Phi(\partial_i\Phi\partial_j\partial_k\Phi+\partial_j\Phi\partial_k\partial_i\Phi+\partial_k\Phi\partial_i\partial_j\Phi)\Big)\Big|^2\nonumber\\
		&+v\Big|\Big(\Phi^2\partial_i\partial_j\partial_k\Phi+2\partial_i\Phi\partial_j\Phi\partial_k\Phi-\Phi(\partial_i\Phi\partial_j\partial_k\Phi+\partial_j\Phi\partial_k\partial_i\Phi+\partial_k\Phi\partial_i\partial_j\Phi)\Big)\Big|^2\nonumber\\
		&+t\Big|\Phi^2\partial_i\partial_j\partial_j\Phi+2\partial_i\Phi\partial_j\Phi\partial_j\Phi-\Phi(\partial_i\Phi\partial_j\partial_j\Phi+\partial_j\Phi\partial_j\partial_i\Phi\partial_j\Phi\partial_i\partial_j\Phi)\Big|^2-m^2\Phi^*\Phi\label{2}
	\end{align}
	Now let us concentrate only on the term that yields the $v$ part of the lagrangian ,
	\begin{equation}
		\mathcal{L}^v_{ijk}(\Phi)=\Phi^2\partial_i\partial_j\partial_k\Phi+2\partial_i\Phi\partial_j\Phi\partial_k\Phi-\Phi(\partial_i\Phi\partial_j\partial_k\Phi+\partial_j\Phi\partial_k\partial_i\Phi+\partial_k\Phi\partial_i\partial_j\Phi)\label{3}
	\end{equation}
	which was also derived here in (\ref{m22}).\\
	Under a general local transformation,
	\begin{equation}
		\Phi \to \Phi' = e^{iw(x)}\Phi\label{4}
	\end{equation}
	(\ref{3}) transforms as,
	\begin{equation}
		\mathcal{L}^v_{ijk}(\Phi) \to \tilde{\mathcal{L}}^v_{ijk}(\Phi)= e^{3iw(x)}(\mathcal{L}^v_{ijk}(\Phi)+i\Phi^3\partial_i\partial_j\partial_kw)\label{5}
	\end{equation}
	So (\ref{3}) does not transform covariantly under the local gauge transformation (\ref{4}). However, it is evident if $w$ depends quadratically on position $x_i$, as given in (\ref{1}), $|\mathcal{L}^v_{ijk}|^2$ will be invariant.\\\\
	Under a general local transformation, covariance can be achieved by adding an extra term in $\mathcal{L}^v_{ijk}(\Phi)$ itself, which also transforms appropriately along with (\ref{4}) under local transformation,
	\begin{equation}
		\mathcal{L}^v_{ijk}(\Phi,A_{ijk})=(\mathcal{L}^v_{ijk}(\Phi)-i\Phi^3A_{ijk})\quad \longrightarrow\quad e^{3iw(x)}(\mathcal{L}^v_{ijk}(\Phi)-i\Phi^3A_{ijk})\label{6}
	\end{equation}
	where $A_{ijk}$ is a rank-3 symmetric tensor gauge field and should transform under local transformation as 
	\begin{equation}
		A_{ijk} \to A'_{ijk}=A_{ijk}+\partial_i\partial_j\partial_k w\label{a13}
	\end{equation}
	to preserve the covariance of the lagrangian. So the total interacting lagrangian is now given by 
	\begin{equation}
		\mathcal{L}^v=|\mathcal{L}^v_{ijk}|^2-i\Big(\Phi^3\mathcal{L}_{ijk}^v(\Phi^*)-{\Phi^*}^3\mathcal{L}^v_{ijk}(\Phi)\Big)A_{ijk}+|\Phi|^6A_{ijk}^2\label{s1}
	\end{equation}
	Now at this point we again refer to section-2 where we have shown that for lagrangian having quadratic shift symmetry, the vector current and the fully symmetric rank-3 tensor current can be related as  (\ref{d27}),
	\begin{equation*}
		j_i=\partial_j\partial_kj_{ijk}
	\end{equation*}
	Now coupling the vector current with a vector gauge field will give
	\begin{equation}
		A_i  j_i=A_i(\partial_j\partial_kj_{ijk})=\frac{1}{3}\Big(\partial_i\partial_jA_k+\partial_j\partial_kA_i+\partial_k\partial_iA_j\Big)j_{ijk}=:-A_{ijk}j_{ijk}
	\end{equation}
	which enables us to identify
	\begin{equation}
		A_{ijk}=-\frac{1}{3}\Big(\partial_i\partial_jA_k+\partial_j\partial_kA_i+\partial_k\partial_iA_j\Big)\label{a12}
	\end{equation}
	One important point to note here is that, $A_{ijk}$ can also be recast in terms of rank-2 tensor gauge fields. The coupling between the rank-2 tensor current and rank-2 tensor gauge field can alternatively be written as
	\begin{equation}
		A_{ij}j_{ij}=A_{ij}(\partial_kj_{ijk})=-\frac{1}{3}(\partial_kA_{ij}+\partial_jA_{ik}+\partial_iA_{kj})j_{ijk}=-A_{ijk}j_{ijk}\label{196}
	\end{equation}
	which enables us to identify,
	\begin{equation}
		A_{ijk}=\frac{1}{3}\Big(\partial_iA_{jk}+\partial_jA_{ik}+\partial_kA_{ij}\Big)=-\frac{1}{3}\Big(\partial_i\partial_jA_k+\partial_j\partial_kA_i+\partial_k\partial_iA_j\Big)\label{mm1}
	\end{equation}
	One can check easily using the gauge transformation rule of vector field $A_i$'s (\ref{k26}) or $A_{ij}$ (\ref{a2}), that $A_{ijk}$ (\ref{a12},\ref{mm1}) indeed transforms as (\ref{a13}), which is consistent with our construction.\\\\
	Now the minimal gauging prescription with $A_i$ is a generalisation of that discussed in the earlier sub-section where a rank-2 tensor field was required.
	As happened earlier, we show that the non-standard gauging (\ref{6}) using a rank-3 tensor field is rephrased in favour of the standard minimal prescription using a vector field. The first step is to symmetrize the triple and double derivatives in (\ref{2}) as  shown below,
	\begin{align}
		\mathcal{L}_{ijk}^v(\Phi)=&\Phi^2\partial_i\partial_j\partial_k\Phi+2\partial_i\Phi\partial_j\Phi\partial_k\Phi-\Phi(\partial_i\Phi\partial_j\partial_k\Phi+\partial_j\Phi\partial_k\partial_i\Phi+\partial_k\Phi\partial_i\partial_j\Phi)\nonumber\\
		=&\frac{1}{3}\Phi^2(\partial_i\partial_j\partial_k\Phi+\partial_j\partial_k\partial_i\Phi+\partial_k\partial_i\partial_j \Phi)+2\partial_i\Phi\partial_j\Phi\partial_k\Phi\nonumber\\
		&-\frac{1}{2}\Phi(\partial_i\Phi\partial_j\partial_k\Phi+\partial_i\Phi\partial_k\partial_j\Phi+\partial_j \Phi\partial_k\partial_i\Phi+\partial_j\Phi\partial_i\partial_k\Phi+\partial_k\Phi\partial_i\partial_j\Phi+\partial_k\Phi\partial_j\partial_i\Phi) \label{7}
	\end{align}
	Next, minimal prescription is imposed by replacing ordinary derivatives $\partial_i$ by their covariant expressions $(\partial_i+iA_i)$, where $A_i$ is the usual vector gauge field. Then (\ref{7}) will take the form
	\begin {align}
	&\mathcal{L}_{ijk}^v(\Phi,A)=\frac{1}{3}\Phi^2\Big((\partial_i+iA_i)(\partial_j+iA_j)(\partial_k+iA_k)\Phi+....\Big)
	+2(\partial_i+iA_i)\Phi(\partial_j+iA_j)\Phi(\partial_k+iA_k)\Phi\nonumber\\
	&-\frac{1}{2}\Phi\Big((\partial_i+iA_i)\Phi(\partial_j+iA_j)(\partial_k+iA_k)\Phi+...
	+(\partial_k+iA_k)\Phi(\partial_j+iA_j)(\partial_i+iA_i)\Phi\Big)\label{8}
\end{align}
A series of remarkable simplifications reduces this to,
\begin{equation}
	\mathcal{L}_{ijk}^v(\Phi,A)=\mathcal{L}_{ijk}^v(\Phi)
	+\frac{i}{3}\Phi^3\Big(\partial_i\partial_jA_k+\partial_j\partial_kA_i+\partial_k\partial_iA_j\Big)\label{9}
\end{equation}
So we have reproduced the same result as shown in (\ref{6}) using standard minimal coupling, if one identifies $A_{ijk}$ as (\ref{a12}).\\ \\
The important point is that the gauging prescription in terms of a tensor field, given in the literature is exactly reproducible by the standard minimal prescription $\partial_i \,\to\, \partial_i+iA_i$ with an appropriate symmetrization of the higher derivatives. On top of that there is an additional freedom of developing the theory in terms of a rank-2 gauge field. It is therefore instructive to carry out a detailed analysis of the full interacting theory in all these formulations. As we show, some nice features are uncovered.
\subsection{Gauge coupling from iterative Noether prescription}
We now discuss the explicit realization of the currents for quadratic shift symmetric lagrangians, directly using Noether's prescription. So let us choose the $v$ part of the lagrangian (\ref{2}).The conserved currents are defined analogous to (\ref{d25}), which are given by
\begin{align}
	J^i&=wj^i-(\partial_jw)j^{ij}+(\partial_j\partial_kw)j^{jki}=\frac{\partial\mathcal{L}_0}{\partial(\partial_i\Phi)}\delta\Phi+\frac{\partial\mathcal{L}_0}{\partial(\partial_i\partial_j\Phi)}\partial_j(\delta\Phi)-\Big(\partial_j\frac{\partial\mathcal{L}_0}{\partial(\partial_i\partial_j\Phi)}\Big)\delta\Phi\nonumber\\
	&+\partial_j\partial_k\Big(\frac{\partial\mathcal{L}_0}{\partial(\partial_i\partial_j\partial_k\Phi)}\Big)\delta\Phi-\partial_j\Big(\frac{\partial\mathcal{L}_0}{\partial(\partial_i\partial_j\partial_k\Phi)}\Big)\partial_k(\delta\Phi)+\frac{\partial\mathcal{L}_0}{\partial(\partial_i\partial_j\partial_k\Phi)}\partial_j\partial_k(\delta\Phi)+c.c\label{hh1}
\end{align}
Equating the coefficients of $c,c_i,c_{ij}$ on opening up $w$, we find,
\begin{equation}
	j^i=\partial_jj_{ij}=\partial_j\partial_kj_{ijk}=i\partial_j\partial_k\Big(\Phi^3\mathcal{L}^v_{ijk}(\Phi^*)-{\Phi^*}^3\mathcal{L}^v_{ijk}(\Phi)\Big)\label{h14}
\end{equation}
where $\mathcal{L}^v_{ijk}(\Phi)$ is defined in (\ref{3}) and $\mathcal{L}^v_{ijk}(\Phi^*)$ is its complex conjugate.\\\\
We now try to adopt the iterative Noether's prescription to obtain the gauge coupling for the above theory. According to this scheme the first correction to the lagrangian is given by,
\begin{equation}
	\mathcal{L}_1=j^iA_i=i\Big(\Phi^3\mathcal{L}^v_{ijk}(\Phi^*)-{\Phi^*}^3\mathcal{L}^v_{ijk}(\Phi)\Big)\partial_k\partial_jA_i\label{h15}
\end{equation}
As can be seen from the above expression, the derivative on the matter fields still persists in $\mathcal{L}^v_{ijk}(\Phi)$ and its complex conjugate. So we expect yet another correction to the lagrangian, which is obtained by first finding the correction to the current,
\begin{equation}
	j^i=\frac{2}{3}\partial_k\partial_j\Big(|\Phi|^6(\partial_i\partial_jA_k+\partial_j\partial_kA_i+\partial_i\partial_kA_j)\Big)\label{h16}
\end{equation}
where $\mathcal{L}_1$ (\ref{h15}) is used in (\ref{hh1})  to get this current. So the contribution to the lagrangian is now given by,
\begin{equation}
	\mathcal{L}_2=\frac{1}{9}|\Phi|^6(\partial_i\partial_jA_k+\partial_j\partial_kA_i+\partial_i\partial_kA_j)^2\label{h17}      
\end{equation}
so that the current is reproduced by (\ref{h6}).
The iteration method terminates here as there is no further derivatives on $\Phi$. So the total gauged lagrangian is given by,
\begin{align}
	\mathcal{L}&=\Big|\mathcal{L}_{ijk}^v|^2
	+\frac{i}{3}\Big(\Phi^3\mathcal{L}^v_{ijk}(\Phi^*)-{\Phi^*}^3\mathcal{L}^v_{ijk}(\Phi)\Big)(\partial_i\partial_jA_k+\partial_j\partial_iA_j+\partial_k\partial_jA_i)\nonumber\\
	&+\frac{1}{9}|\Phi|^6(\partial_i\partial_jA_k+\partial_j\partial_kA_i+\partial_i\partial_kA_j)^2\label{h19}
\end{align}
We can also recast the above analysis in terms of the tensor field defined in (\ref{a22}). Then (\ref{h15}) can be re-expressed as,
\begin{equation}
	\mathcal{L}_1=-\frac{i}{3}\Big(\Phi^3\mathcal{L}^v_{ijk}(\Phi^*)-{\Phi^*}^3\mathcal{L}^v_{ijk}(\Phi)\Big)(\partial_iA_{jk}+\partial_jA_{ik}+\partial_kA_{ij})=j_{ij}A_{ij}\label{h20}
\end{equation}
As $\mathcal{L}_1$ consists derivatives of $\phi$, we can derive the current again  using $\mathcal{L}_1$ in (\ref{hh1}) and the new (rank-2 tensor) current is given by
\begin{equation}
	j_{ij}=-\frac{2}{3}\partial_k\Big(|\Phi|^6(\partial_iA_{jk}+\partial_jA_{ik}+\partial_kA_{ij})\Big)\label{h21}
\end{equation}
and the new contribution to the lagrangian is now given by
\begin{equation}
	\mathcal{L}_2=\frac{1}{9}|\Phi|^6(\partial_iA_{jk}+\partial_jA_{ik}+\partial_kA_{ij})^2\label{h22}
\end{equation}
so that the current (\ref{h21}) is given by the definition (\ref{h11}). Hence the final lagrangian takes the form
\begin{equation}
	\mathcal{L}=|\mathcal{L}^v_{ijk}|^2-\frac{i}{3}\Big(\Phi^3\mathcal{L}^v_{ijk}(\Phi^*)-{\Phi^*}^3\mathcal{L}^v_{ijk}(\Phi)\Big)(\partial_iA_{jk}+\partial_jA_{ik}+\partial_kA_{ij})+\frac{1}{9}|\Phi|^6(\partial_iA_{jk}+\partial_jA_{ik}+\partial_kA_{ij})^2\label{h23}
\end{equation}
Now let us proceed to realise the above theory in terms of rank-3 tensor fields. The lagrangian (\ref{h20}) can be re-expressed as
\begin{equation}
	\mathcal{L}_1=-i\Big(\Phi^3\mathcal{L}^v_{ijk}(\Phi^*)-{\Phi^*}^3\mathcal{L}^v_{ijk}(\Phi)\Big)A_{ijk}=-j_{ijk}A_{ijk}\label{h24}
\end{equation}
where this coupling of $j^{ijk}$ to $A_{ijk}$ is defined in (\ref{196}). From this lagrangian we get a new rank-3 tensor current as
\begin{equation}
	j_{ijk}=-2|\Phi|^6A_{ijk}\label{h25}
\end{equation}
The contribution to the lagrangian is as follows,
\begin{equation}
	\mathcal{L}_2=|\Phi|^6A_{ijk}^2\label{h26}
\end{equation}
so that the current (\ref{h25}) can be reproduced by
\begin{equation}
	j^{ijk}=-\frac{\delta}{\delta A_{ijk}}\int \mathcal{L}_2
\end{equation}
Thus the total lagrangian can be written as
\begin{equation}
	\mathcal{L}=|\mathcal{L}^v_{ijk}|^2-i\Big(\Phi^3\mathcal{L}^v_{ijk}(\Phi^*)-{\Phi^*}^3\mathcal{L}^v_{ijk}(\Phi)\Big)A_{ijk}+|\Phi|^6A_{ijk}^2\label{h27}
\end{equation}
Here we have shown that how the theory described by (\ref{s1}) in the section-5 can be derived from a scalar theory coupled with vector gauge fields using the relationship between various bootstrapped currents or equivalently the various gauge fields.

\subsection{Full interacting theory and equations of motion in (3+1) dimension}
In this section we shall derive the equations of motion of quadratic shift symmetric lagrangian in 3+1 dimension and try to establish a connection between such equations of  various gauge fields such as $A_i,A_{ij},A_{ijk}$ which are interrelated with each other as shown in section-5, although the relations were established at only algebraic level. With this dynamical treatment we can establish that the relations (\ref{a12},\ref{mm1}) shown in previous sections are robust in nature.\\\\
\textbf{Theory with rank-3 tensor gauge fields}\\
We have already shown in section-5, that for a quadratic shift symmetric lagrangian the gauging prescription can be carried out with a rank-3 symmetric tensor gauge field $A_{ijk}$. We shall now make the gauge fields dynamic and calculate the equations of motion for all the fields. For that let us first define the electric and magnetic fields as\footnote{Note that as both the magnetic fields of rank-3 and rank-2 tensor theories are tensors of rank-2, we have used $\tilde{B}$ and $B$ to distinguish. $\tilde{B}$ is magnetic field for rank-3 theory and $B$ is the magnetic field for rank-2 theory.}, 
\begin{align}
	E_{ijk}&=\partial_0A_{ijk}+\partial_i\partial_j\partial_kA_0\nonumber\\
	\tilde{B}_{ij}&=\partial_k\partial_m(\partial_iA_{jkm}-\partial_jA_{ikm})\label{g1}
\end{align}
The pure gauge field lagrangian can be written as
\begin{align}
	\mathcal{L}_0&=\frac{1}{2}(E_{ijk}^2-\tilde{B}_{ij}^2)\nonumber\\
	&=\frac{1}{2}\Big[(\partial_i\partial_j\partial_kA_0)^2+(\partial_0A_{ijk})^2+2(\partial_0A_{ijk})(\partial_i\partial_j\partial_kA_0)\Big]-\frac{1}{2}\Big[\partial_m\partial_k(\partial_iA_{jkm}-\partial_jA_{ikm})\Big]^2\label{g6}
\end{align}
So the total lagrangian is written as
\begin{align}
	\mathcal{L}&=\mathcal{L}_0+\frac{i}{2}\Phi^*\overleftrightarrow{\partial}_0\Phi-\Phi^*\Phi A_0+ \Big|\mathcal{L}^v_{ijk}(\Phi)-i\Phi^3A_{ijk}\Big|^2\\
	&= \mathcal{L}_0+\frac{i}{2}\Phi^*\overleftrightarrow{\partial}_0\Phi-\Phi^*\Phi A_0+ \Big|\mathcal{L}^v_{ijk}(\Phi)\Big|^2-j_{ijk}A_{ijk}
\end{align}
where $\mathcal{L}^v_{ijk}$ is defined in (\ref{3}) and tensor current $j_{ijk}$ is given by
\begin{equation}
	j_{ijk}=i\Big(\Phi^3\mathcal{L}^v_{ijk}(\Phi^*)-{\Phi^*}^3\mathcal{L}^v_{ijk}(\Phi)\Big)+|\Phi|^6A_{ijk}
\end{equation}
Now the equation of motion of $A_0,A_{ijk}$ are, respectively, given by
\begin{align}
	\partial_i\partial_j\partial_kE_{ijk}&=-\Phi^*\Phi\label{g2}\\
	\partial_0E_{ijk}-\frac{1}{3}\partial_l\Big(\partial_i\partial_j\tilde{B}_{lk}+\partial_j\partial_k\tilde{B}_{li}+\partial_k\partial_i\tilde{B}_{lj}\Big)&=-j_{ijk}\label{g3}
\end{align}
The canonical momentum corresponding to $A_{ijk}$ is given by
\begin{equation}
	\pi_{ijk}=\frac{\partial \mathcal{L}}{\partial(\partial^i\partial^j\partial^k\Phi)}=E_{ijk}\label{g11}
\end{equation}
and the algebra between the basic fields is defined as,
\begin{align}
	&\{A_{ijk}(\vec{x}),E_{mnp}(\vec{y})\}=\frac{1}{6}[\delta_{im}(\delta_{jn}\delta_{kp}+\delta_{jp}\delta_{kn})+\delta_{jn}(\delta_{im}\delta_{kp}+\delta_{ip}\delta_{km})+\delta_{kp}(\delta_{im}\delta_{jn}+\delta_{in}\delta_{jm})]\delta(\vec{x}-\vec{y})\nonumber\\
	& \{A_{ijk}(\vec{x}),A_{mnl}(\vec{y})\}=0,\qquad\{\pi_{ijk}(\vec{x}),\pi_{mnl}(\vec{y})\}=0\label{g19}
\end{align}
The corresponding canonical Hamiltonian can be written as,
\begin{equation}
	H_c=\frac{1}{2}(E_{ijk}^2+\tilde{B}_{ij}^2)+...\label{g12}
\end{equation}
Using the Poission algebra (\ref{g19}) we can find out the Hamiltonian equation of motion as   
\begin{equation}
	\partial_0\tilde{B}_{ij}+\frac{2}{3}\partial_n\partial_k(\partial_iE_{jnk}-\partial_jE_{ink})=0\label{g55}
\end{equation}
and the other Hamiltonian equation involving time derivative of $E_{ijk}$ is same as that of (\ref{g3}).\\\\
\textbf{Theory with rank-2 tensor gauge fields}\\\\
We shall now use the map (\ref{mm1}) between rank-3 gauge field and rank-2 gauge field to write down the gauge field lagrangian in terms of rank-2 tensor gauge fields. Using (\ref{mm1}) the lagrangian can be recast as
\begin{equation}
	\mathcal{L}_0=(\partial_i\partial_j\partial_kA_0)^2+\frac{1}{3}(\partial_i\partial_0A_{jk})^2+2(\partial_i\partial_0A_{jk})(\partial_i\partial_j\partial_kA_0)+\frac{2}{3}(\partial_i\partial_0A_{jk})(\partial_j\partial_0A_{ik})-\frac{4}{9}[\partial^2\partial_m(\partial_iA_{jm}-\partial_jA_{im})]^2\label{g7}
\end{equation}
The electric ($E_{ij}$) and magnetic ($B_{ij}$) fields are given by (\ref{a8},\ref{a9}). The total lagrangian is written as
\begin{equation}
	\mathcal{L}=\mathcal{L}_0+\frac{i}{2}\Phi^*\overleftrightarrow{\partial}_0\Phi-\Phi^*\Phi A_0+ \Big|\mathcal{L}^v_{ijk}(\Phi)\Big|^2-j_{ij}A_{ij}
\end{equation}
where $j_{ij}=\partial_kj_{ijk}$. The equations of motion of $A_0,A_{ij}$ are, respectively, given by
\begin{align}
	\partial^2\partial_j\partial_kE_{jk}=-\Phi^*\Phi\label{g10}\\
	\frac{1}{3}\partial_0\Big(\partial^2E_{ij}+\partial_i\partial_kE_{jk}+\partial_j\partial_kE_{ik}\Big)+\frac{2}{9}\partial^4\partial_k(\partial_iB_{kj}+\partial_jB_{ki})=-j_{ij}\label{g10a}
\end{align}
The canonical momentum conjugate to $A_{ij}$ is, \begin{equation}
	\pi_{ij}=\frac{\partial \mathcal{L}}{\partial(\partial^i\partial^j\Phi)}=\frac{1}{3}(\partial^2E_{ij}+\partial_k\partial_iE_{jk}+\partial_k\partial_jE_{ik})=\partial_kE_{ijk}\label{g13}
\end{equation}
satisfying the following Poission algebra
\begin{equation}
	\{A_{ij}(x),\pi_{kl}(y)\}=\frac{1}{2}(\delta_{ik}\delta_{jl}+\delta_{il}\delta_{jk})\delta(x-y),\quad \{A_{ij}(x),A_{kl}(y)\}=0=\{\pi_{ij}(x),\pi_{kl}(y)\}\label{g18}
\end{equation}
The corresponding canonical Hamiltonian is given by
\begin{equation}
	H_c=\frac{1}{2}\Big(\frac{1}{\partial^2}\partial_k\pi_{ij}\Big)^2+\frac{2}{9}(\partial^2B_{ij})^2+...\label{g15}
\end{equation}
Using (\ref{g18}) one can find out the Hamiltonian equation of motion as
\begin{equation}
	\partial_0B_{ij}-\frac{2}{3}\partial_k(\partial_iE_{jk}-\partial_jE_{ik})=0\label{g30}
\end{equation}\\
\vspace{5pt}
\textbf{Equation of motion for usual vector gauge fields}\\\\
Now we shall use the map (\ref{a12}) to write down the lagrangian (\ref{g6}) in terms of usual vector fields $A_i$. The corresponding electric $E_i$ and magnetic $B_i$ fields follow their usual definitions. The gauge lagrangian is given by
\begin{align}
	\mathcal{L}_{0}&=\frac{1}{2}\Big[(\partial_i\partial_j\partial_kA_0)^2+\frac{1}{9}(\partial_0\partial_i\partial_jA_k+\partial_0\partial_i\partial_kA_j+\partial_0\partial_k\partial_jA_i)^2\nonumber\\
	&-\frac{2}{3}\partial_0(\partial_i\partial_jA_k+\partial_i\partial_kA_j+\partial_k\partial_jA_i)\Big]-\frac{1}{18}(\partial^4\epsilon_{ijk}\epsilon_{klm}\partial_lA_m)^2\label{g20}
\end{align}
so that the total lagrangian can be written as
\begin{equation}
	\mathcal{L}=\mathcal{L}_{0}+\frac{i}{2}\Phi^*\overleftrightarrow{\partial}_0\Phi-\Phi^*\Phi A_0+ \Big|\mathcal{L}^v_{ijk}(\Phi)\Big|^2-j_{i}A_{i}
\end{equation}
The equations of motion of $A_0, A_i$ are given by
\begin{align}
	\partial^4 \partial_iE_i&=-\Phi^*\Phi\label{g4}\\
	\frac{\partial^4}{3}\Big[\partial_0E_i-\frac{\partial^4}{3}\epsilon_{ijk}\partial_jB_k\Big]+\frac{2}{3}\partial^2\partial_0\partial_i(\partial.E)&=-j_i\label{g5}    
\end{align}
The canonical momentum conjugate to $A_i$ is given by \begin{equation}
	\pi_i=\frac{\partial \mathcal{L}}{\partial\dot{A}^i}=\frac{1}{3}\partial^2(\partial^2\delta_{ij}+2\partial_i\partial_j)E_j\label{g14}
\end{equation}
which satisfies the algebra
\begin{equation}
	\{A_{i}(\vec{x}),\pi_j(\vec{y})\}=\delta_{ij}\delta(\vec{x}-\vec{y}),\quad\{A_i(\vec{x}),A_j(\vec{y})\}=0=\{\pi_i(\vec{x}),\pi(\vec{y})\}
\end{equation}
while the canonical hamiltonian is obtained by a Legendre transformation from (\ref{g20}),
\begin{equation}
	H_c=\frac{1}{2}\Big(\frac{\partial_j\partial_k\pi_i}{\partial^4}\Big)^2+\frac{1}{2}\Big(\frac{1}{3}\epsilon_{ijk}\partial^4B_k\Big)^2\label{g17}
\end{equation}
which yields the equation of motion as
\begin{equation}
	\partial_0B_i+\frac{1}{3}\epsilon_{mni}(\partial_mE_n-\partial_nE_m)=0\label{m20}
\end{equation}
\textbf{Comparative study}\\\\
We now use the relations between various gauge fields and electric fields to find out the mapping between the equations in various theories discussed above. First we use the mapping between rank-3 tensor fields to rank-2 tensor fields given by (\ref{mm1}). The corresponding mapping between the electric and magnetic fields are given by
\begin{equation}
	E_{ijk}=\frac{1}{3}(\partial_iE_{jk}+\partial_jE_{ik}+\partial_kE_{ij})\label{g8}
\end{equation}
\begin{equation}
	\tilde{B}_{ij}=\partial_k\partial_m(\partial_iA_{jkm}-\partial_jA_{ikm})=\frac{2}{3}\partial^2\partial_m(\partial_iA_{jm}-\partial_jA_{im})=-\frac{2}{3}\partial^2B_{ij}\label{g9}   
\end{equation}
Now using (\ref{g8},\ref{g9}) in (\ref{g2}) and (\ref{g3}), we get
\begin{align}
	\partial^2\partial_j\partial_kE_{jk}&=-\Phi^*\Phi\label{m11}\\
	\frac{1}{3}\partial_0(\partial_iE_{jk}+\partial_jE_{ik}+\partial_kE_{ij})+\frac{2\partial^2}{9}\partial_l(\partial_i\partial_jB_{lk}+\partial_j\partial_kB_{li}+\partial_k\partial_iB_{lj})&=-j_{ijk}\label{mm12}
\end{align}
Acting $\partial_k$ on (\ref{mm12}), we get (\ref{g10a}) which is the equation of motion of $E_{ij}$ for rank-2 tensorial theory.\\
The canonical momenta of the respective theories are given by
\begin{equation}
	\pi_{ijk}=E_{ijk}=\frac{1}{\partial^2}\partial_k\pi_{ij} 
\end{equation}
Exploiting this result the hamiltonian (\ref{g12}) exactly maps to (\ref{g15}). Further, using the maps (\ref{g8},\ref{g9}), equation (\ref{g55}) can be recast in the following form 
\begin{equation}
	\partial^2\Big[\partial_0B_{ij}-\frac{2}{3}\partial_k(\partial_iE_{jk}-\partial_jE_{ik})\Big]=0
\end{equation}
which is compatible with the Hamiltonian equation for rank-2 tensor fields given by (\ref{g30}).\\\\
Now we use the map (\ref{a12}) from rank-3 tensor gauge fields to rank-1 vector gauge fields to write down the following relations between the corresponding electric and magnetic fields as
\begin{align}
	E_{ijk}=& \frac{1}{3}(\partial_i\partial_jE_k+\partial_j\partial_kE_i+\partial_k\partial_iE_j)\label{mm13}\\
	\tilde{B}_{ij}=&-\frac{1}{3}\partial^4\epsilon_{ijk}B_k\label{mm14}
\end{align}
Using these mapping the lagrangian equations of motion for rank-3 tensor fields (\ref{g2},\ref{g3}) can be mapped to usual vector field equations as
\begin{align}
	\partial^4(\partial.E)&=-\Phi^*\Phi\label{mm15}\\
	\frac{\partial_0}{3}[\partial_i\partial_jE_k+\partial_j\partial_kE_i+\partial_k\partial_iE_j]+\frac{\partial^4\partial_l}{9}[\partial_i\partial_j\epsilon_{lkm}+\partial_j\partial_k\epsilon_{lim}+\partial_k\partial_i\epsilon_{ljm}]B_m&=-j_{ijk}
	\label{mm16}
\end{align}
By acting $\partial_j\partial_k$ on (\ref{mm16}), we get (\ref{g5}), the equation of motion of $E_{i}$ of vector theory, while the Gauss law is exactly reproduced (\ref{g5},\ref{mm15}).\\
The canonical momenta corresponding to the respective theories can be mapped as
\begin{equation}
	\pi_{ijk}=\frac{\partial_j\partial_k\pi_i}{\partial^4} \label{mm17}
\end{equation}
This relation is used to show that the Hamiltonian (\ref{g12}) can exactly be mapped to the hamiltonian of vector theory (\ref{g17}). The Hamiltonian equation of motion (\ref{g55}) is then mapped to the following equation using (\ref{mm13},\ref{mm14}),
\begin{equation}
	\partial^2(\partial_0B_i+\frac{1}{3}\epsilon_{mni}(\partial_mE_n-\partial_nE_m))=0
\end{equation}
the solution of which is exactly (\ref{m20}).
The comparison can be clearly understood using the following chart:\\
\begin{tabular}{|m{1.4cm}|m{5.5cm}|m{5.7cm}|m{4.5cm}|}
	\hline
	&Gauge field&Electric field&Magnetic field\\
	\hline
Fields&$A_i,\,A_{ij},\,A_{ijk}$&$E_i=\partial_iA_0-\partial_0A_i,\,E_{ij},\,E_{ijk}$&$B_i=\epsilon_{ijk}\partial_jA_k,B_{ij},\tilde{B}_{ij}$\\
	\hline
Derived maps&$A_{ijk}=\frac{1}{3}\Big(\partial_iA_{jk}+\partial_jA_{ik}+\partial_kA_{ij}\Big)$&$E_{ijk}=\frac{1}{3}\Big(\partial_iE_{jk}+\partial_jE_{ik}+\partial_kE_{ij}\Big)$&$\tilde{B}_{ij}=-\frac{2}{3}\partial^2B_{ij}$\\
	\hline
&$A_{ijk}=-\frac{1}{3}\Big(\partial_i\partial_jA_k+\partial_j\partial_kA_i+\partial_k\partial_iA_j\Big)$&$E_{ijk}=\frac{1}{3}\Big(\partial_i\partial_jE_k+\partial_j\partial_kE_i+\partial_k\partial_iE_j\Big)$&$\tilde{B}_{ij}=-\frac{1}{3}\partial^4\epsilon_{ijk}B_k$\\
	\hline
\end{tabular}
\begin{center}
	Table V : Comparison of various fields in (3+1) dimension for quadratic shift symmetric theory
	\end{center}
The above table shows the relationship of rank-3 tensor fields ($A_{ijk},E_{ijk},\tilde{B}_{ij}$) with rank-2 tensor fields ($A_{ij},E_{ij},B_{ij}$) and with vector fields $(A_i,E_i,B_i)$.\\
\begin{small}
	\begin{tabular}{ | m{1.5cm} | m{3.6cm}| m{5.5cm} |m{5.5cm}|} 
		\hline
		&Gauss Law & Faraday's Law & Ampere's Law \\
		\hline
		Rank-3 tensor theory & $\partial_i\partial_j\partial_kE_{ijk}=-\Phi^*\Phi$ & $\partial_0E_{ijk}-\frac{1}{3}\partial_l\Big(\partial_i\partial_j\tilde{B}_{lk}+\partial_j\partial_k\tilde{B}_{li}+\partial_k\partial_i\tilde{B}_{lj}\Big)=-j_{ijk}$& $\partial_0\tilde{B}_{ij}+\frac{2}{3}\partial_n\partial_k(\partial_iE_{jnk}-\partial_jE_{ink})=0$\\ 
		\hline
	     
		Using map (\ref{g8},\ref{g9}) in rank-3 theory &$\partial^2\partial_j\partial_kE_{jk}=-\Phi^*\Phi$ 
		& $\frac{1}{3}\partial_0(\partial_iE_{jk}+\partial_jE_{ik}+\partial_kE_{ij})+\frac{2\partial^2}{9}\partial_l(\partial_i\partial_jB_{lk}+\partial_j\partial_kB_{li}+\partial_k\partial_iB_{lj})=-j_{ijk}$ &$\partial^2\Big[\partial_0B_{ij}-\frac{2}{3}\partial_k(\partial_iE_{jk}-\partial_jE_{ik})\Big]=0$\\
		\hline
		&&Acting $\partial_k$ on the above equation& Acting $\frac{1}{\partial^2}$ on the above equation\\
		\hline
			Rank-2 tensor theory& $\partial^2\partial_j\partial_kE_{jk}=-\Phi^*\Phi$&  $\frac{1}{3}\partial_0\Big(\partial^2E_{ij}+\partial_i\partial_kE_{jk}+\partial_j\partial_kE_{ik}\Big)+\frac{2}{9}\partial^4\partial_k(\partial_iB_{kj}+\partial_jB_{ki})=-j_{ij}$ &$\partial_0B_{ij}-\frac{2}{3}\partial_k(\partial_iE_{jk}-\partial_jE_{ik})=0$ \\ 
		\hline 
		\hline
		Using map (\ref{mm13},\ref{mm14}) in rank-3 theory & $\partial^4 \partial_iE_i=-\Phi^*\Phi$ & $ \frac{\partial_0}{3}[\partial_i\partial_jE_k+\partial_j\partial_kE_i+\partial_k\partial_iE_j]+\frac{\partial^4\partial_l}{9}[\partial_i\partial_j\epsilon_{lkm}+\partial_j\partial_k\epsilon_{lim}+\partial_k\partial_i\epsilon_{ljm}]B_m=-j_{ijk}$ &   $\partial^2(\partial_0B_i+\frac{1}{3}\epsilon_{mni}(\partial_mE_n-\partial_nE_m))=0 $\\
		\hline
	&	&Acting $\partial_j\partial_k$ on above equation& Acting $\frac{1}{\partial^2}$ on above equation\\
	\hline
			Vector theory& $\partial^4 \partial_iE_i=-\Phi^*\Phi$&$
		\frac{\partial^4}{3}\Big[\partial_0E_i-\frac{\partial^4}{3}\epsilon_{ijk}\partial_jB_k\Big]+\frac{2}{3}\partial^2\partial_0\partial_i(\partial.E)=-j_i$&$  \partial_0B_i+\frac{1}{3}\epsilon_{mni}(\partial_mE_n-\partial_nE_m)=0$ \\ 
		\hline   
	\end{tabular}
\end{small}\\
\begin{center}
	Table VI : Comparison of various equations of motion in (3+1) dimension for quadratic shift
	symmetric theory
\end{center}
The first, fourth and last row represent the equations of motion for theories with rank-3,rank-2 and vector couplings respectively. It is also shown using the relations of table V, that the rank-3 tensor equations are mapped to rank-2 and rank-1 tensor equations. It is either mapped directly (Gauss law) or by using some appropriate derivative operators (Faraday's and Ampere's law).
\section{Conclusion}
Study of polynomial space dependent shift symmetric theories introduced in \cite{griffin,griffin1}, is appealing and is vastly studied recently, especially due to their n-pole conservation laws apart from usual charge conservation stemming from standard invariance under global (constant) transformations.\\
In this paper using a generalised Noether type prescription, we have given a
definition yielding all the n-pole currents. Also, various conservation laws associated with these currents were derived at one stroke using a master equation. It was shown that various tensor currents are intertwined with each other and were related by the action of a tower of spatial derivatives.\\\\
Using this result it was possible to establish a duality between tensor gauge fields with different ranks. Introduction of interactions was the precise mechanism by which these dual relations were derived (see for example, discussion around (\ref{B2})). Thus it was possible to write down the interaction between a gauge field and a current in different ways (using tensors of different ranks) exploiting the duality relations between the currents and the gauge fields.\\
On the other hand, it was shown in \cite{dual}, that using the iterative Noether prescription of Deser \cite{deser} one can introduce matter-gauge interactions in two different ways, the reason of which is in the dual picture stated above. In one case they are introduced by a vector gauge field and the additional terms, besides the free piece, can be written so that the complete theory is the standard minimally coupled theory. Alternatively, one can also express the interactions in terms of a symmetric tensor gauge field recently discovered in fractonic theories. This method of introducing interactions naturally led to a nonstandard ‘minimal prescription’ where the complete theory was expressed by replacing the ordinary double derivatives by additional terms involving tensor fields which was explicitly derived previously in \cite{dual} for linear shift symmetric theory. Symmetrisation of derivatives plays an important role in this prescription. We have expanded this possibility and shown that this non-standard minimal scheme is valid for a general shift symmetric theory. The calculations also indicate that the non-standard minimal coupling prescription is obtained from the standard one by using the map that relates the vector (standard) and tensor (non-standard) gauge fields discussed above.\\
We have considered in details theory of complex scalars interacting either with vector or symmetric tensor gauge fields. It is important to mention that the pure gauge part has two possible options: (a) using the vector gauge fields (b) using the tensor gauge fields. To start with we considered the theory in terms of the tensor fields. Using our maps this was also expressed in favour of the vector fields. Both these theories admit equations of motion (Gauss law, Ampere's law and Faraday's law) which were suitably compared. All computations were done for linear shift symmetric theories in 2+1 and 3+1 dimensions and quadratic shift symmetric theories in 3+1 dimension. 
\section*{Acknowledgement} RB acknowledges support from a (DAE) Raja Ramanna Fellowship (grant no: 1003/(6)/2021/RRF/R$\&$D -II/4031, dated: 20/03/2021). AC acknowledges DST, India for financial support in the form of fellowship. 
\bibliographystyle{unsrt}
\bibliography{main.bib}
\end{document}